%% file: paper.tex
\documentclass{JHEP3}
\usepackage{epsfig} 
 
\input texpref.tex

\newcommand{\be}{\begin{equation}} 
\newcommand{\ee}{\end{equation}} 
\newcommand{\bea}{\begin{eqnarray}} 
\newcommand{\eea}{\end{eqnarray}} 
\newcommand{\bean}{\begin{eqnarray*}} 
\newcommand{\eean}{\end{eqnarray*}} 
\newcommand{\nn}{\nonumber}
\newcommand{\de}{\partial} 
\newcommand{\dd}{\mathrm{d}}

\newcommand{\tr}{\mbox{tr}} 
 
\preprint{MIT-CTP-3646, CERN-PH-TH/2005-084, HUTP-05/A0027} 
 
\title{Gauge Theories from Toric Geometry and Brane Tilings} 
 
\author{\parbox{16cm}{Sebasti\'an Franco$^1$, Amihay Hanany$^1$, Dario Martelli$^2$, 
James Sparks$^3$, David Vegh$^1$, and Brian Wecht$^1$ }\\ 
 
\vspace{0.3 cm}

\parbox[t]{6in} 
{1. Center for Theoretical Physics, 
                   Massachusetts Institute of Technology,\\ 
                 Cambridge, MA 02139, USA.}\\ 
~\\ 
{2. Department of Physics, CERN Theory Division, 
                   1211 Geneva 23, Switzerland.}\\ 
~\\ 
{3. Department of Mathematics, Harvard University, \\ 
                   One Oxford Street, Cambridge, MA 02318, U.S.A.\\ 
                   {\it and} \\ 
                   Jefferson Physical Laboratory, Harvard University, 
                   Cambridge, MA 02138, U.S.A.}\\ 
\vspace{0.5 cm} 
~\\ 
\email{sfranco@mit.edu, hanany@mit.edu, dario.martelli@cern.ch, 
sparks@math.harvard.edu, dvegh@mit.edu, bwecht@mit.edu} }

\abstract{We provide a general set of rules for extracting the data
defining a quiver gauge theory from a given toric Calabi--Yau
singularity. Our method combines information from the
geometry and topology of Sasaki--Einstein manifolds, AdS/CFT, dimers, and 
brane tilings. 
We explain how the field
content, quantum numbers, and superpotential of a superconformal
gauge theory on D3--branes probing a toric Calabi--Yau singularity
can be deduced. The infinite family of toric singularities with
known horizon Sasaki--Einstein manifolds $L^{a,b,c}$ is used to
illustrate these ideas. We construct the corresponding quiver gauge
theories, which may be fully specified by giving a
tiling of the plane by hexagons with certain gluing rules. As
 checks of this construction, we perform $a$-maximisation
as well as $Z$-minimisation to compute the exact R-charges of an
arbitrary such quiver. We also examine a number of examples in
detail, including the infinite subfamily $L^{a,b,a}$, whose smallest
member is the Suspended Pinch Point.} 
 
\begin{document} 
\tableofcontents
 
 
\section{Introduction} 
 
Gauge theories arise within string theory in a variety of different 
ways. One possible approach, which is particularly interesting due 
to its relationship with different branches of geometry, is to use 
D--branes to probe a singularity. The geometry of the singularity 
then determines the amount of supersymmetry, the gauge group 
structure, the matter content and the superpotential interactions on 
the worldvolume of the D--branes. The richest of such examples which 
are both tractable, using current techniques, and also non--trivial, 
are given by the $d=4$ $\mathcal{N}=1$ gauge theories that arise on 
a stack of D3--branes probing a singular Calabi--Yau 3--fold. There 
has been quite remarkable progress over the last year in 
understanding these theories, especially in the case where the 
Calabi--Yau singularity is also toric. In this case one can use 
toric geometry, which by now is an extremely well--developed 
subject, to study the gauge theories. In fact part of this paper is 
devoted to pushing this further and we will show how one can use 
arguments in toric geometry and topology to derive much of the field 
content of a D3--brane probing a toric Calabi--Yau singularity in a 
very simple manner, using essentially only the toric diagram and 
associated gauged linear sigma model. 
 
At low energies the theory on the D3--brane is expected to flow to a 
superconformal fixed point. The AdS/CFT correspondence 
\cite{Maldacena:1997re,Gubser:1998bc,Witten:1998qj} connects the 
strong coupling regime of such gauge theories with supergravity in a 
mildly curved geometry. For the case of D3--branes placed at the 
tips of Calabi--Yau cones over five--dimensional geometries $Y_5$, 
the gravity dual is of the form $AdS_5 \times Y_5$, where $Y_5$ is a 
Sasaki--Einstein manifold 
\cite{Kehagias:1998gn,Klebanov:1998hh,Acharya:1998db,Morrison:1998cs}. 
There has been considerable progress in this subject recently: for a 
long time, there was only one non--trivial Sasaki--Einstein 
five--manifold, $T^{1,1}$, where the metric was known. Thanks to 
recent progress, we now have an infinite family of explicit metrics 
which, when non--singular and simply--connected, have topology $S^2 
\times S^3$. The most general such family is specified by 3 positive 
integers $a,b,c$, with the metrics denoted $L^{a,b,c}$ 
\cite{Cvetic:2005ft,Martelli:2005wy}\footnote{We have changed the notation to $L^{a,b,c}$ to avoid 
confusion with the $p$ and $q$ of $Y^{p,q}$. In our notation, 
$Y^{p,q}$ is $L^{p+q,p-q,p}$.}. When $a=p-q,b=p+q,c=p$ these 
reduce to the $Y^{p,q}$ family of metrics, which have an enhanced 
$SU(2)$ isometry 
\cite{Gauntlett:2004zh,Gauntlett:2004yd,Martelli:2004wu}. Aided by 
the toric description in \cite{Martelli:2004wu}, the entire infinite 
family of gauge theories dual to these metrics was constructed in 
\cite{Benvenuti:2004dy}. These theories have subsequently been 
analysed in considerable detail 
\cite{Herzog:2004tr,Benvenuti:2004wx,Pal:2005mr,Benvenuti:2005wi,Lunin:2005jy, 
Bergman:2005ba,Franco:2005fd,Cascales:2005rj,Burrington:2005zd,Berenstein:2005xa,Franco:2005zu,Benvenuti:2005cz,Bertolini:2005di}. 
There has also been progress on the non--conformal extensions of 
these theories (and others) both from the supergravity 
\cite{Herzog:2004tr,Burrington:2005zd} and gauge theory sides 
\cite{Franco:2005fd,Cascales:2005rj,Berenstein:2005xa,Franco:2005zu,Bertolini:2005di,Franco:2004jz}. 
These extensions exhibit many interesting features, such as cascades 
\cite{Klebanov:2000hb} and dynamical supersymmetry breaking. In 
addition to the $Y^{p,q}$ spaces, there are also several other 
interesting infinite families of geometries which have been studied 
recently: the $X^{p,q}$ spaces \cite{Hanany:2005hq}, deformations of 
geometries with $U(1)\times U(1)$ isometry \cite{Lunin:2005jy}, and 
deformations of geometries with $U(1)^3$ isometry 
\cite{Ahn:2005vc,Gauntlett:2005zz}.

Another key ingredient in obtaining the gauge theories dual to 
singular Calabi--Yau manifolds is the principle of $a$--maximisation 
\cite{Intriligator:2003jj}, which permits the determination of exact 
R--charges of superconformal field theories. Recall that all $d=4$ 
$\mathcal{N}=1$ gauge theories possess a $U(1)_R$ symmetry which is 
part of the superconformal group $SU(2,2|1)$. If this superconformal 
R--symmetry is correctly identified, many properties of the gauge 
theory may be determined. $a$--maximisation 
\cite{Intriligator:2003jj} is a simple procedure -- maximizing a 
cubic function -- that allows one to identify the R--symmetry from 
among the set of global symmetries of any given gauge theory.
Plugging the superconformal R-charges into this cubic function
gives exactly the central charge $a$ of the SCFT 
\cite{Anselmi:1998am,Anselmi:1998ys,Henningson:1998gx}.
Although here we will focus on superconformal theories with known 
geometric duals, $a$--maximization is a general procedure which 
applies to any $\mathcal{N}=1$ $d=4$ superconformal field theory, and 
has been studied in this context in a number of recent works, with 
much emphasis on its utility for proving the $a$--theorem 
\cite{Kutasov:2003iy,Intriligator:2003mi, 
Kutasov:2003ux,Csaki:2004uj,Barnes:2004jj,Kutasov:2004xu,Barnes:2005zn}. 
 
In the case that the gauge theory has a geometric dual, one can use 
the AdS/CFT correspondence to compute the volume of the dual 
Sasaki--Einstein manifold, as well as the volumes of certain 
supersymmetric 3--dimensional submanifolds, from the R--charges. For 
example, remarkable agreement was found for these two computations 
in the case of the $Y^{p,q}$ singularities 
\cite{Bertolini:2004xf,Benvenuti:2004dy}. Moreover, a general 
geometric procedure that allows one to compute the volume of any 
toric Sasaki--Einstein manifold, as well as its toric supersymmetric 
submanifolds, was then given in \cite{Martelli:2005tp}. In 
\cite{Martelli:2005tp} it was shown that one can determine the Reeb 
vector field, which is dual to the R--symmetry, of any toric 
Sasaki--Einstein manifold by minimising a function $Z$ that depends 
only on the toric data that defines the singularity. For example, 
the volumes of the $Y^{p,q}$ manifolds are easily reproduced this 
way. Remarkably, one can also compute the volumes of manifolds for 
which the metric is not known explicitly. In all cases agreement has 
been found between the geometric and field theoretic calculations. 
This was therefore interpreted as a geometric dual of 
$a$--maximisation in \cite{Martelli:2005tp}, although to date there 
is no general proof that the two extremal problems, within the class 
of superconformal gauge theories dual to toric Sasaki--Einstein 
manifolds, are in fact equivalent. 
 
Another important step was achieved recently by the introduction of 
dimer technololgy as a tool for studying $\mathcal{N}=1$ gauge 
theories. Although it has been known in principle how to compute
toric data dual to a given SCFT and vice versa \cite{Feng:2000mi,Feng:2001xr},
these computations are often very computationally expensive, even for
fairly small quivers. Dimer technology greatly simplifies this process,
and turns previously intractable calculations into easily solved problems. 
The initial connection between toric geometry and dimers 
was suggested in \cite{Hanany:2005ve}; the connection to 
$\mathcal{N}=1$ theories was proposed and explored in 
\cite{Franco:2005rj}. A crucial realization that enables one to use 
this tool is that all the data for an $\mathcal{N}=1$ theory can be 
simply represented as a periodic tiling (``brane tiling'') of the 
plane by polygons with an even number of sides:  the faces represent 
gauge groups, the edges represent bifundamentals, and the vertices 
represent superpotential terms. This tiling has a physical meaning 
in Type IIB string theory as an NS5--brane wrapping a holomorphic curve 
(the edges of the tiling) with D5--branes (the faces) ending on the 
NS5--brane. The fact that the polygons have an even number of sides 
is equivalent to the requirement that the theories be anomaly free; 
by choosing an appropriate periodicity, one can color the vertices 
of such a graph with two colours (say, black and white) so that a 
black vertex is adjacent only to white vertices. Edges that stretch 
between black and white nodes are the dimers, and there is a simple 
prescription for computing a weighted adjacency matrix (the Kasteleyn 
matrix) which gives the partition function for a given graph. This 
partition function encodes the toric diagram for the dual geometry 
in a simple way, thus enabling one to have access to many properties 
of both the gauge theory and the geometry. The tiling construction 
of a gauge theory is a much more compact way of describing the 
theory than the process of specifying both a quiver and the 
superpotential, and we will use this newfound simplicitly rather 
extensively in this paper. 
 
In this paper, we will use this recent progress in geometry, field 
theory, and dimer models to obtain a lot of information about
gauge theories dual to general toric Calabi-Yau cones. Our 
geometrical knowledge will specify many requirements of the gauge 
theory, and we describe how one can read off gauge theory quantities
rather straightforwardly from the geometry. As a particular example
of our methods, we construct the gauge theories dual to the
recently discovered $L^{a,b,c}$ geometries.
We will realize the geometrically derived requirements 
by using the brane tiling approach. Since the $L^{a,b,c}$ 
spaces are substantially 
more complicated than the $Y^{p,q}$'s, we will not give a closed 
form expression for the gauge theory. We will, however, specify all 
the necessary building blocks for the brane tiling, and discuss how 
these building blocks are related to quantities derived from the 
geometry. 

The plan of the paper is as follows: In Section 2, we 
discuss how to read gauge theory data from a given toric geometry. 
In particular, we give a detailed prescription for computing the 
quantum numbers ({\it e.g.} baryon charges, flavour charges, and 
R--charges) and multiplicities for the different fields in the gauge 
theory. Section 3 applies these results to the $L^{a,b,c}$ 
spaces. We derive the toric diagram for a 
general $L^{a,b,c}$ geometry, and briefly review the metrics 
\cite{Cvetic:2005ft,Martelli:2005wy} for these theories. We compute 
the volumes of the supersymmetric 3--cycles in these 
Sasaki--Einstein spaces, and discuss the constraints these put on 
the gauge theories. In Section 4 we discuss how our geometrical 
computations constrain the superpotential, and describe how one may 
always find a phase of the gauge theory with at most only three 
different types of interactions.
In Section 5, we 
prove that 
$a$--maximisation reduces to the same equations required by the 
geometry for computing R-charges and central charges. Thus we show
that $a$--maximisation and the geometric computation agree.
In Section 6, we construct 
the gauge theories dual to the $L^{a,b,c}$ spaces by using the brane 
tiling perspective, and give several examples of interesting 
theories. In particular, we describe a particularly simple infinite 
subclass of theories, the $L^{a,b,a}$ theories, for which we can 
simply specify the toric data and brane tiling. We check via 
$Z$--minimisation and $a$--maximisation that all volumes and 
dimensions reproduce the results expected from AdS/CFT. Finally,
in the Appendix, we give some more interesting examples which
use our construction.
 
\bigskip

{\bf Note:} While this paper was being finalised, we were made aware of other work in \cite{Benvenuti:2005ja}, 
which has some overlap with our results. Similar conclusions have been reached in \cite{Butti:2005sw}.

\section{Quiver content from toric geometry} 
 
\label{section_quiver_geometry} 
 
In this section we explain how one can extract a considerable amount of 
information about the gauge theories on D3--branes probing 
toric Calabi--Yau singularities using simple geometric 
methods. In particular, we show that there is always a 
distinguished set of fields whose multiplicities, baryon charges, 
and flavour charges can be computed straightforwardly using the toric data. 
 
\subsection{General geometrical set--up} 
 
Let us first review the basic geometrical set--up. For more 
details, the reader is referred to \cite{Martelli:2005tp}. 
Let $(X,\omega)$ be a toric Calabi--Yau cone of complex dimension $n$, 
where $\omega$ is the K\"ahler form on $X$. In particular 
$X=C(Y)\cong\IR^+\times Y$ has an isometry group containing an 
$n$--torus, $T^n$. A conical metric on $X$ which is both Ricci--flat 
and K\"ahler then gives a Sasaki--Einstein metric on the base of the 
cone, $Y$. The moment map for the torus 
action 
exhibits $X$ as a Lagrangian $T^n$ fibration over a 
strictly convex rational polyhedral cone $\mathcal{C}\subset\R^n$. 
This is a subset of $\R^n$ of the form 
\be\label{conehead} 
\mathcal{C} = \{y\in\R^n\mid (y,v_A)\geq 0, A=1,\ldots,D\}~.\ee 
Thus $\mathcal{C}$ is made by intersecting $D$ hyperplanes through the 
origin in order to make a convex polyhedral cone. 
Here $y\in\R^n$ are coordinates on $\IR^n$ and $v_A$ are the inward 
pointing normal vectors to the $D$ hyperplanes, or \emph{facets}, 
that define the 
polyhedral cone. The normals are rational and hence one can normalise 
them to be primitive\footnote{A vector $v\in\IZ^n$ is primitive 
if it cannot be written as $mv^{\prime}$ with 
$v^{\prime}\in\IZ^n$ and $\IZ\ni m>1$.} 
elements of the lattice $\IZ^n$. We also assume this set of vectors is 
minimal in the sense that removing any vector $v_A$ in 
the definition (\ref{conehead}) changes $\mathcal{C}$. 
The condition that $\mathcal{C}$ be \emph{strictly} convex is simply 
the condition that it is a cone over a convex polytope. 
 
\begin{figure}[ht] 
  \epsfxsize = 7cm 
  \centerline{\epsfbox{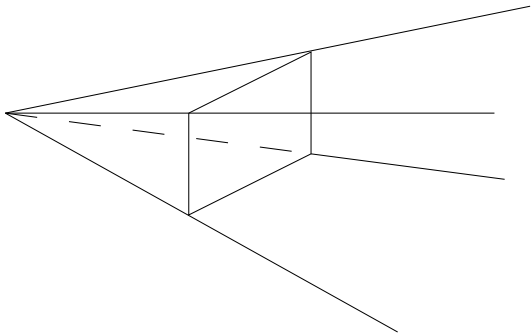}} 
  \caption{A four--faceted polyhedral cone in $\IR^3$.} 
  \label{polycone} 
\end{figure} 
 
The condition that $X$ is Calabi--Yau, $c_1(X)=0$, implies 
that the vectors $v_A$ may, by an appropriate 
$SL(n;\IZ)$ transformation of the torus, be all written as $v_A=(1,w_A)$. 
In particular, in complex dimension $n=3$ we may therefore represent any 
toric Calabi--Yau cone by a convex lattice polytope in 
$\IZ^2$, where the vertices are simply the vectors $w_A$. 
This is usually called the toric diagram. 
 
From the vectors $v_A$ one can reconstruct $X$ 
as a K\"ahler quotient or, more physically, as the 
classical vacuum moduli space of a gauged linear sigma model (GLSM). 
To explain this, denote by $\Lambda \subset \IZ^n$ the span 
of the normals $\{v_A\}$ over $\IZ$. This is a lattice of maximal rank 
since $\mathcal{C}$ is strictly convex. 
Consider the linear map 
\begin{eqnarray}\label{linear} 
A: &&\IR^D \rightarrow \IR^n \nn \\ 
& & \ e_A\mapsto v_A\end{eqnarray} 
which maps each standard orthonormal basis 
vector $e_A$ of $\IR^D$ to the vector $v_A$. 
This induces a map of tori 
\be 
T^D\cong \IR^D/\IZ^D \rightarrow \IR^n/\Lambda~.\ee 
In general the 
kernel of this map is 
$\mathcal{A}\cong T^{D-n}\times \Gamma$ where $\Gamma$ is a finite 
abelian group. Then $X$ is given by the K\"ahler quotient 
\be\label{sq} 
X = \IC^D//\mathcal{A}~.\ee 
Recall we may write this more explicitly as follows. The 
torus $T^{D-n}\subset T^D$ is specified by a 
charge matrix $Q_I^A$ with integer coefficients, $I=1,\ldots,D-n$, and we define 
\be 
\mathcal{K} \equiv \left\{(Z_1,\ldots,Z_D)\in\IC^D\mid\sum_{A}Q_{I}^A 
|Z_A|^2 = 0\right\}\subset \IC^D\ee 
where $Z_A$ denote complex coordinates on $\IC^D$. In GLSM language, 
$\mathcal{K}$ is simply the space of solutions to the D--term 
equations. Dividing out by gauge transformations gives the quotient 
\be X = \mathcal{K}/ T^{D-n}\times\Gamma~.\ee We also denote by $L$ 
the link of $\mathcal{K}$ with the sphere $S^{2D-1}\subset \IC^D$. 
We then have a fibration \be\label{fibration} 
\mathcal{A}\hookrightarrow L \rightarrow Y\ee where $Y$ is the 
Sasakian manifold which is the base of the cone $X=C(Y)$. For a 
general set of vectors $v_A$, the space $Y$ will not be smooth. In 
fact typically one has orbifold singularities. $Y$ is smooth if and 
only if the polyhedral cone is \emph{good} \cite{L}, although we 
will not enter into the general details of this here -- see, for 
example, \cite{Martelli:2005tp}. 
 
Finally in this subsection we note some topological 
properties of $Y$, in the case that $Y$ is a smooth manifold. 
In \cite{Ltop} it is shown that $L$ has trivial 
homotopy groups in dimensions $0$, $1$ and $2$. 
From the long  exact homotopy sequence for the fibration 
(\ref{fibration}) 
one concludes that \cite{Ltop} 
\bea\label{pis} 
&&\pi_1(Y) \cong  \pi_0(\mathcal{A})\cong \Gamma\cong\IZ^n/\Lambda\\ \nn 
&&\pi_2(Y) \cong  \pi_1(\mathcal{A})\cong\IZ^{D-n}~.\eea 
In particular, $Y$ is simply--connected 
if and only if the $\{v_A\}$ span $\IZ^n$ over $\IZ$. In fact 
we will assume this throughout this paper -- any finite quotient 
of a toric singularity will correspond to an orbifold of the 
corresponding gauge theory, and this process is well--understood by 
now. 
 
From now on we also restrict to the physical case of complex 
dimension $n=3$. Moreover, throughout this section we assume that 
the Sasaki--Einstein manifold $Y$ is smooth. The reason for this 
assumption is firstly to simplify the geometrical and topological 
analysis, and secondly because the physics in the case that $Y$ is 
an orbifold which is \emph{not} a global quotient of a smooth 
manifold is not well--understood. However, as we shall see later, 
one can apparently relax this assumption with the results 
essentially going through without modification. The various 
cohomology groups that we introduce would then need replacing by 
their appropriate orbifold versions.

\subsection{Quantum numbers of fields} 
 
In this subsection we explain how one can deduce the quantum numbers 
for a certain distinguished set of fields in any toric quiver gauge 
theory. Recall that, quite generally, $N$ D3--branes placed at a 
toric Calabi--Yau singularity have an AdS/CFT dual that may be 
described by a toric quiver gauge theory. In particular, the matter 
content is specified by giving the number of gauge groups, $N_g$, 
and number of fields $N_f$, together with the charge assignments of 
the fields. In fact these fields are always bifundamentals (or 
adjoints). This means that the matter content may be neatly 
summarised by a quiver diagram. 
 
We may describe the toric singularity 
as a convex lattice polytope in $\IZ^2$ or 
by giving the GLSM charges, as described 
in the previous section. 
By setting each complex coordinate $Z_A=0$, $A=1,\ldots,D$, 
one obtains a toric divisor $D_A$ in the Calabi--Yau cone. 
This is also a cone, with $D_A=C(\Sigma_A)$ where $\Sigma_A$ is a 
3--dimensional supersymmetric submanifold of $Y$. Thus in 
particular wrapping a D3--brane over $\Sigma_A$ gives rise 
to a BPS state, which via the AdS/CFT correspondence is conjectured 
to be dual to a dibaryonic operator in the dual gauge theory. 
We claim that there is always a distinguished subset of the 
fields, for any toric quiver gauge theory, which are associated 
to these dibaryonic states. To explain this, recall 
that given any bifundamental field $X$, one can construct the 
dibaryonic operator 
\bea\label{dib} 
\mathcal{B}[X] = \epsilon^{\alpha_1\ldots\alpha_N} X_{\alpha_1}^{\beta_1} 
\ldots X_{\alpha_N}^{\beta_N} \epsilon_{\beta_1\ldots\beta_N}\eea 
using the epsilon tensors of the corresponding two $SU(N)$ 
gauge groups. This is dual to a D3--brane wrapped on a supersymmetric 
submanifold, for example one of the $\Sigma_A$. In fact to 
each toric divisor $\Sigma_A$ let us associate a bifundamental field 
$X_A$ whose corresponding dibaryonic operator (\ref{dib}) is 
dual to a D3--brane wrapped on $\Sigma_A$. These 
fields in fact have multiplicities, as we explain momentarily. 
In particular each field in such a multiplet has the same 
baryon charge, flavour charge, and R--charge.

\subsubsection{Multiplicities} 
 
Recall that $D_A=\{Z_A=0\}=C(\Sigma_A)$ where $\Sigma_A$ is 
a 3--submanifold of $Y$. To each such submanifold we 
associated a bifundamental field $X_A$. As we now explain, these 
fields have multiplicities given by the simple formula 
\bea 
\label{mul} 
m_A = |(v_{A-1},v_A,v_{A+1})| 
\eea 
where we have defined the cyclic identification $v_{D+1}=v_1$, and 
we list the normal vectors $v_A$ in order around the polyhedral cone, 
or equivalently, the toric diagram. Here $(\cdot,\cdot,\cdot)$ denotes 
a $3\times3$ determinant, as in \cite{Martelli:2005tp}. 
 
In fact, when $Y$ is smooth, each 
$\Sigma_A$ is a Lens space $L(n_1,n_2)$ for appropriate 
$n_1$ and $n_2$. To see this, note that each $\Sigma_A$ is a 
principle $T^2$ fibration over an interval, say $[0,1]$. 
By an $SL(2;\IZ)$ transformation one can always arrange that 
at $0$ the $(1,0)$--cycle collapses, and 
at $1$ the $(n_1,n_2)$--cycle collapses. It is well--known that 
this can be equivalently described as the quotient of 
$S^3\subset\IC^2$ by the $\IZ_{n_1}$ action 
\be 
(z_1,z_2)\rightarrow (z_1\omega_{n_1},z_2\omega_{n_1}^{n_2})\ee 
where $\mathrm{hcf}(n_1,n_2)=1$ and $\omega_{n_1}$ denotes 
an $n_1$th root of unity. These spaces have a rich 
history, and even the classification of homeomorphism types 
is rather involved. We shall only need to know that 
$\pi_1(L(n_1,n_2))\cong\IZ_{n_1}$, which is immediate from the 
second definition above. 
 
Consider now wrapping a D3--brane over some smooth $\Sigma$, where 
$\pi_1(\Sigma)=\IZ_{m}$. As we just explained, when 
$\Sigma$ is toric, it is necessarily some Lens space $L(m,n_2)$. 
In fact, when the order of the fundamental group is greater than 
one, there is not a single such 
D3--brane, but in fact $m$ D3--branes. The reason is that, for each $m$, 
we can turn on a flat line bundle for the $U(1)$ gauge field on 
the D3--brane worldvolume. Indeed, recall that line bundles 
on $\Sigma$ are classified topologically by $H^2(\Sigma;\IZ)\cong 
H_1(\Sigma;\IZ)\cong \pi_1(\Sigma)$, where the last 
relation follows for abelian fundamental group. A torsion line bundle always 
admits a flat connection, which has zero energy. Since these 
D3--brane states have different charge -- namely torsion D1--brane charge -- 
they must correspond to different operators in the gauge theory. 
However, as will become clear, these operators all have the same baryon 
charge, flavour charge and R--charge. 
We thus learn that the multiplicity of the bifundamental field 
$X_A$ associated with $D_A$ is given by $m$. 
 
It remains then to relate $m$ to the formula (\ref{mul}). 
Without loss of generality, pick a facet $A$, and suppose 
that the normal vector is $v_A=(1,0,0)$. The facet is itself a 
polyhedral cone in the $\IR^2$ plane transverse to this vector. 
To obtain the normals that define this cone we simply 
project $v_{A-1}$, $v_{A+1}$ onto the plane. Again, by 
a special linear transformation we may take these 2--vectors 
to be $(0,1)$, $(n_1,-n_2)$, respectively, for some integers $n_1$ and $n_2$. 
One can then verify that this toric diagram indeed corresponds to the cone over 
$L(n_1,n_2)$, as defined above. By direct calculation we now see that 
\be 
|(v_{A-1},v_A,v_{A+1})|= |(0,1)\times(n_1,-n_2)| = n_1 
\ee 
which is the order of $\pi_1(\Sigma_A)$. The 
determinant is independent of the choice of basis we have made, and 
thus this relation is true in general, thus proving the formula 
(\ref{mul}). One can verify this formula in a large number of examples 
where the gauge theories are already known.

\subsubsection{Baryon charges} 
 
In this subsection we explain how one can deduce the baryonic 
charges of the fields $X_A$. Recall that, in general, the toric 
Sasaki--Einstein manifold $Y$ arises from a quotient by a torus \bea 
T^{D-3}\hookrightarrow L \rightarrow Y~.\eea This fibration can be 
thought of as $D-3$ circle fibrations over $Y$ with total space $L$. 
Equivalently we can think of these as complex line bundles 
$\mathcal{M}_I$. Let $C_I$, $I=1,\ldots,D-3$, denote the Poincar\'e 
duals of the first Chern classes of these bundles. Thus they are 
classes in $H_3(Y;\IZ)$. Recall from (\ref{pis}) that $\pi_2(Y)\cong 
\IZ^{D-3}$ when $Y$ is smooth. Provided $Y$ is also 
simply--connected\footnote{Recall this is also one of our 
assumptions in this section.} one can use the Hurewicz isomorphism, 
Poincar\'e duality and the universal coefficients theorem to deduce 
that \bea H_3(Y;\IZ)\cong\IZ^{D-3}~.\eea In particular note that the 
number of independent 3--cycles is just $D-3$. A fairly 
straightforward calculation\footnote{For example, one can use the 
Gysin sequence for each circle in turn.} in algebraic topology shows 
that the classes $C_I$ above actually generate the homology group 
$H_3(Y;\IZ)\cong\IZ^{D-3}$. Thus $\{C_I\}$ form a basis of 3--cycles 
on $Y$.

In Type IIB supergravity 
one can Kaluza--Klein reduce the Ramond--Ramond four--form potential 
$C_4$ to obtain $D-3$ gauge fields $A_I$ in the $AdS_5$ space: 
\be 
C_4 = \sum_{I=1}^{D-3} A_I \wedge\mathcal{H}_I~.\ee 
Here $\mathcal{H}_I$ is a harmonic 3--form on $Y$ that is Poincar\'e dual 
to the 3--cycle 
$C_I$. In the superconformal gauge theory, which recall may be thought of 
as living on the conformal boundary of $AdS_5$, these become $D-3$ global 
$U(1)$ symmetries 
\be\label{charge} 
U(1)_B^{D-3}~.\ee 
These are baryonic symmetries precisely because the D3--brane is charged under 
$C_4$ and a D3--brane wrapped over a supersymmetric submanifold of 
$Y$ is interpreted as a dibaryonic state in the gauge theory. 
Indeed, the $\Sigma_A$ are precisely such a set of 
submanifolds. 
 
Again, a fairly standard calculation in toric 
geometry then shows that topologically 
\be\label{toprel} 
[\Sigma_A] = \sum_{I=1}^{D-3} Q_I^A C_I\in H_3(Y;\IZ)~.\ee 
This perhaps requires a little explanation. Each 
GLSM field $Z_A$, $A=1,\ldots,D$, can be viewed as a 
section of a complex line bundle $\mathcal{L}_A$ over $Y$. They are 
necessarily sections of line bundles, rather than functions, 
because the fields $Z_A$ are charged under the 
torus $T^{D-3}$. Now $Z_A=0$ is the zero section of the line 
bundle associated to $Z_A$, and by definition this cuts out the 
submanifold $\Sigma_A$ on $Y$. Moreover, the first Chern class 
of this line bundle is then Poincar\'e dual to $[\Sigma_A]$. 
Recall that the charge matrix $Q$ specifies the 
embedding of the torus $T^{D-3}$ in $T^D$, which then acts on the 
fields/coordinates $Z_A$; the element $Q_I^A$ specifies 
the charge of $Z_A$, which is a section of $\mathcal{L}_A$, 
under the circle $\mathcal{M}_I$. This means that the two sets of 
line bundles are related by 
\bea 
\mathcal{L}_A = \bigotimes_{I=1}^{D-3} \mathcal{M}_I^{Q_I^A}~.\eea 
Taking the first Chern class of this relation and applying 
Poincar\'e duality then proves (\ref{toprel}). 
 
It follows that the baryon charges of the fields $X_A$ are given 
precisely 
by the matrix $Q$ that enters in defining the GLSM. 
Thus if $B_I[X_A]$ denotes the baryon charge of $X_A$ under 
the $I$th copy 
of $U(1)$ in (\ref{charge}) we have 
\be 
B_I [X_A] = Q_I^A~.\ee 
Note that from the Calabi--Yau condition the charges of the linear 
sigma model sum to zero \bea \sum_A B_I [X_A] = \sum_A Q_I^A= 
0\qquad \quad I=1, \dots , D-3~. \eea Moreover, the statement that 
\be \sum_A v_A^i [\Sigma_A] =0\ee may then be interpreted as saying 
that, for each $i$, one can construct a state in the gauge theory of 
zero baryon charge by using $v_A^i$ copies of the field $X_A$, for 
each $A$. 
 
\subsubsection{Flavour charges} 
 
In this subsection we explain how one can compute the flavour 
charges of the $X_A$. Recall that the horizon Sasaki--Einstein 
manifolds have at least a $U(1)^3$ isometry since they are toric. By 
definition a flavour symmetry in the gauge theory is a non--R--symmetry --  
that is, the supercharges are left invariant under such 
a symmetry. The geometric dual of this statement is that the Killing 
spinor $\psi$ on the Sasaki--Einstein manifold $Y$ is left invariant 
by the corresponding isometry. Thus a Killing vector field $V_F$ is 
dual to a flavour symmetry in the gauge theory if and only if \be 
\mathcal{L}_{V_F} \psi = 0\label{bobby}\ee where $\psi$ is a Killing 
spinor on $Y$. In fact there is always precisely a $U(1)^2$ subgroup 
of $U(1)^3$ that satisfies this condition. This can be shown by 
considering the holomorphic $(3,0)$ form of the corresponding 
Calabi--Yau cone \cite{Martelli:2005tp}. It is well known that this is 
constructed from the Killing spinors as a bilinear \be \Omega  = 
\psi^c \, \Gamma_{(3)}\, \psi~, \ee where $\Gamma_{(3)}$ is the 
totally antisymmetrised product of $3$ gamma matrices in 
Cliff$(6,0)$. In particular, in the basis in which the normal 
vectors of the polyhedral cone $\mathcal{C}$ are of the form 
$v_A=(1,w_A)$, the Lie algebra elements $(0,1,0),(0,0,1)$ generate 
the group $U(1)^2_F$ of flavour isometries. Note that, for 
$Y^{p,q}$, one of these $U(1)_F$ symmetries is enhanced to an 
$SU(2)$ flavour symmetry. However, $U(1)_F^2$ is the generic case. 
 
We would like to determine the charges of the fields $X_A$ under 
$U(1)^2_F$. In fact in the gauge theory this symmetry group is 
far from unique -- one is always free to mix any flavour symmetry 
with part of the baryonic symmetry group $U(1)^{D-3}_B$. The baryonic 
symmetries are distinguished by the fact that mesons in the gauge theory, 
for example constructed from closed loops in a 
quiver gauge theory, should have zero baryonic charge. Thus the 
flavour symmetry group is unique only up to mixing with 
baryonic symmetries, and of course mixing with each other. 
 
This mixing ambiguity has a beautiful geometric interpretation. Recall 
that 
the Calabi--Yau cone $X$ is constructed as a symplectic quotient 
\be 
X=\IC^D//T^{D-3}\ee 
where the torus $T^{D-3}\subset T^D$ is defined by the kernel of the map 
\bea 
A:&& \R^D\rightarrow \R^3\\ 
&& e_A\mapsto v_A~.\eea 
More precisely the kernel of $A$ is generated by the 
matrix $Q_I^A$, which in turn 
defines a sublattice $\Upsilon$ of $\IZ^D$ of rank $D-3$. 
The torus is then $T^{D-3}=\IR^{D-3}/\Upsilon$. 
We may also consider the 
quotient $\IZ^D/\Upsilon$. The map induced from $A$ then maps this 
quotient space isomorphically onto $\IZ^3$ 
and the corresponding torus $T^3=T^D/ T^{D-3}$ is then precisely the 
torus isometry of $X$. 
 
Let us pick two elements $\alpha_1,\alpha_2$ of 
$\IZ^D$ that map to the basis vectors 
$(0,1,0),(0,0,1)$ under $A$. From the last paragraph these are defined 
only up to elements of the lattice $\Upsilon$, and thus may be 
considered as elements of the quotient $\IZ^D/\Upsilon$. 
Geometrically, $\alpha_1, \alpha_2$ define circle subgroups 
of $T^D$ that descend to the two $U(1)$ flavour isometries 
generated by $(0,1,0)$ and $(0,0,1)$. The charges of the complex 
coordinates $Z_A$ on $\IC^D$ are then simply $\alpha_1^A$, $\alpha_2^A$ for 
each $A=1,\ldots,D$. However, as discussed in the last subsection, the $Z_A$ 
descend to complex line bundles on $Y$ whose Poincar\'e duals are 
precisely the submanifolds $\Sigma_A$. Thus the flavour charges 
of $X_A$ may be identified with $\alpha_1^A$, $\alpha_2^A$. 
Moreover, 
by construction, each $\alpha$ was unique only up to 
addition by some element in the lattice $\Upsilon$ generated by 
$Q_I^A$. But as we just saw in the previous subsection, this 
is precisely the set of baryon charges in the gauge theory. We thus 
see that the ambiguity in the choice of flavour symmetries in the 
gauge theory is in 1--1 correspondence with the ambiguity in choosing 
$\alpha_1,\alpha_2$.

\subsubsection{R--charges} 
 
The R--charges were treated in reference \cite{Martelli:2005tp}, so we will be 
brief here. Let us begin by emphasising that all the quantities 
computed so far can be extracted in a simple way from the toric 
data, or equivalently from the charges of the gauged linear sigma 
model, without the need of an explicit metric. In \cite{Martelli:2005tp}, it was 
shown that the total volume of any toric Sasaki--Einstein manifold, 
as well as the volumes of its supersymmetric toric submanifolds, can 
be computed by solving a simple extremal problem which is defined in 
terms of the polyhedral cone $\mathcal{C}$. This toric data is 
encoded in a function $Z$, which depends on a ``trial'' Reeb vector 
living in $\IR^3$. Minimising $Z$ determines the Reeb vector for the 
Sasaki--Einstein metric on $Y$ uniquely, and as a result one can 
compute the volumes of the $\Sigma_A$. This is a geometric analogue 
of $a$--maximisation \cite{Intriligator:2003jj}. Indeed, recall that the volumes are 
related to the R--charges of the corresponding fields $X_A$ by the 
simple formula \bea R[X_A] & = & \frac{\pi}{3} \frac{\vol 
(\Sigma_A)}{\vol (Y)}~. \eea 
This formula has been used in many AdS/CFT calculations to compare
the R--charges of dibaryons with their corresponding 3--manifolds
\cite{Berenstein:2002ke,Intriligator:2003wr,Herzog:2003wt,Herzog:2003dj}.

Moreover, in \cite{Martelli:2005tp} a general 
formula relating the volume of supersymmetric submanifolds to the 
total volume of the toric Sasaki--Einstein manifold was given. This 
reads \bea \pi \sum_{A=1}^{D} \vol(\Sigma_A) = 6 \, \vol (Y)~. 
\label{msyvolumes} \eea Then the physical interpretation of 
(\ref{msyvolumes}) is that the R--charges of the bifundamental fields 
$X_A$ sum to 2: \bea \sum_A R[X_A] & = & 2~. \label{W=2} \eea This 
is related to the fact that each term in the superpotential is 
necessarily the sum \bea \sum_{A=1}^D \Sigma_A\eea and the 
superpotential has R--charge 2 by definition. We shall discuss this 
further in Section 4. 
 

\section{The $L^{a,b,c}$ toric singularities} 
\label{section_Labc} 
  
In the remainder of this paper we will be interested in 
the specific GLSM with charges 
\bea 
Q=(a,-c,b,-d)
\label{lsmgen} 
\eea 
where of course $d=a+b-c$ in order to satisfy the 
Calabi--Yau condition. We will define this singularity to be $L^{a,b,c}$. 
The reason we choose this family is two--fold: firstly, the 
Sasaki--Einstein metrics are known explicitly 
in this case \cite{Cvetic:2005ft,Martelli:2005wy} and, secondly, this family is 
sufficiently simple that we will be able to give a general 
prescription for constructing the gauge theories. 
  
Let us begin by noting that this is essentially the most 
general GLSM with four charges, and hence the most general 
toric quiver gauge theories with a single $U(1)_B$ symmetry, 
up to orbifolding. Indeed, provided all the charges are 
non--zero, either two have the same sign or else three have the 
same sign. The latter are in fact just orbifolds of 
$S^5$, and this case where all but one of the charges have the 
same sign is slightly degenerate. Specifically, the 
charges $(e,f,g,-e-f-g)$ describe 
the orbifold of $S^5\subset \IC^3$ by $\IZ_{e+f+g}$ with weights 
$(e,f,g)$. The polyhedral cones therefore have \emph{three} 
facets, and not four, or equivalently the $(p,q)$ web has 
3 external legs. By our general analysis there is therefore 
no $U(1)$ baryonic symmetry, as expected. Indeed, note that 
setting $Z_4=0$ does \emph{not} give a divisor in this case, 
since the remaining charges are all positive and there is no 
solultion to the remaining D--terms. The Sasaki--Einstein metrics 
are just the quotients of the round metric on $S^5$ and these 
theories are therefore not particularly interesting. 
In the case that one of the charges is zero, we instead obtain 
$\mathcal{N}=2$ orbifolds of $S^5$, which are also well--studied. 
  
We are therefore left with the case that two charges have the 
same sign. In (\ref{lsmgen}) we therefore take all integers to be positive. 
Without loss of generality we may of course take 
$0<a\leq b$. Also, by swapping $c$ and $d$ if necessary, we can 
always arrange that $c\leq b$. By definition 
$\mathrm{hcf}(a,b,c,d)=1$ in order that the $U(1)$ action 
specified by (\ref{lsmgen}) is effective, and it then follows 
that any three integers are coprime. 
The explicit Sasaki--Einstein metrics on the horizons of these 
singularities were constructed in \cite{Cvetic:2005ft}. The toric 
description above was then given in \cite{Martelli:2005wy}. 
The manifolds were named 
$L^{p,q,r}$ in reference \cite{Cvetic:2005ft} but, following \cite{Martelli:2005wy}, we have 
renamed these $L^{a,b,c}$ in order to avoid confusion with 
$Y^{p,q}$.  Indeed, notice that these spaces reduce to $Y^{p,q}$ 
when $c=d=p$, and then 
$a=p-q$, $b=p+q$. In particular there is an enhanced $SU(2)$ 
symmetry in the metric in this limit. 
It is straightforward to determine when the space 
$Y=L^{a,b,c}$ is non--singular: 
each of the pair $a,b$ must be coprime to each of 
$c,d$. This condition is necessary to avoid codimension 
\emph{four} orbifold singularities on $Y$. 
To see this,  consider setting $Z_1=Z_4=0$. If $b$ and $c$ had 
a common factor $h$, then the circle action specified by (\ref{lsmgen}) 
would factor through a cyclic group $\IZ_h$ of order $h$, and this would 
descend to a local orbifold group on the quotient space. In fact it 
is simple to see that this subspace is just an $S^1$ family of 
$\IZ_h$ orbifold singularities. All such singularities arise 
in this way. 
When $Y=L^{a,b,c}$ is non--singular it follows from the last section 
that $\pi_2(Y)\cong \IZ$ and hence $H_2(Y;\IZ)\cong\IZ$. By Smale's 
theorem $Y$ is therefore diffeomorphic to $S^2\times S^3$. 
In particular there is one 3--cycle and hence one 
$U(1)_B$ for these theories. 
  
\begin{figure}[ht] 
  \epsfxsize = 5cm 
  \centerline{\epsfbox{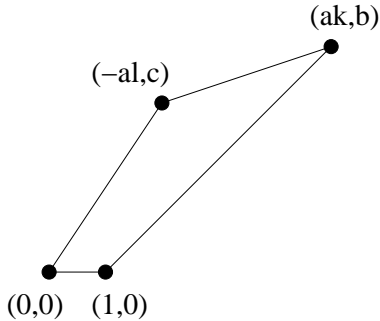}} 
  \caption{Toric diagram for the $L^{a,b,c}$ geometries.}
  \label{toric_generic}
\end{figure}

The toric diagram can be described by an appropriate 
set of four vectors $v_A=(1,w_A)$. 
We take the following set 
\bea 
w_1 = [1,0]\qquad w_2  = [ak,b]\qquad w_3 =[-al,c]\qquad w_4 = [0,0] 
\eea 
where $k$ and $l$ are two integers satisfying 
\bea 
c \, k + b \, l =1 
\eea 
and we have assumed for simplicity of exposition 
that $\mathrm{hcf}(b,c)=1$. This toric diagram is depicted in \fref{toric_generic}.

 The 
solution to the above equation always 
exists by Euclid's algorithm. Moreover, there is a countable 
infinity of solutions to this equation, where one shifts $k$ and $l$ 
by $-tb$ and $tc$, respectively, for any integer $t$. However, it is simple to check that 
different solutions are related by the $SL(2;\IZ)$ transformation 
\bea\label{icecream} 
\left(\begin{array}{cc} 1 & -ta\\ 0& 1\end{array}\right)\eea 
acting on the $w_A$, as must be the case of course. 
The kernel of the linear map 
(\ref{linear}) is then generated by the charge vector $Q$ in (\ref{lsmgen}). 
  
It is now simple to see that the toric diagram for $L^{a,b,c}$ always admits a 
triangulation with $a+b$ triangles. It is well known that this gives 
the number of 
gauge groups $N_g$ in the gauge theory. To see this one uses the fact 
that the area of the toric diagram is the Euler number of 
the (any) completely resolved Calabi--Yau $\tilde{X}$ obtained 
by toric crepant resolution, and then for 
toric manifolds this is the dimension of the even cohomology 
of $\tilde{X}$. Now on $0$, $2$ and $4$--cycles in $\tilde{X}$ one can wrap 
space--filling D3, D5 and D7--branes, respectively, and these 
then form a basis of fractional branes. The gauge groups may then 
be viewed as the gauge groups on these fractional branes. 
By varying the K\"ahler moduli of $\tilde{X}$ one can blow down 
to the conical singularity $X$. The holomorphic part of the gauge theory 
is independent of the K\"ahler moduli, which is why the matter 
content of the superconformal 
gauge theory can be computed at large volume in this way. 
To summarise, we have 
\bea 
N_g = a+b~.\eea 
Note that, different from $Y^{p,q}$, the number of gauge 
groups for $L^{a,b,c}$ can be \emph{odd}. 
  
\begin{figure}[ht] 
  \epsfxsize = 8cm 
  \centerline{\epsfbox{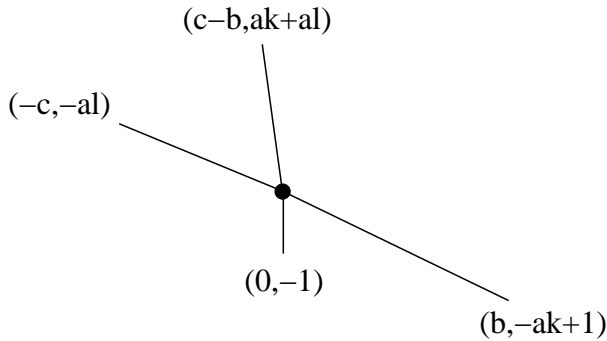}} 
  \caption{(p,q)-web for the $L^{a,b,c}$ theories.}
  \label{pq_generic}
\end{figure}

We may now draw the  $(p,q)$ web \cite{Aharony:1997bh,Aharony:1997ju,Leung:1997tw}. Recall that this is simply the 
graph--theoreric dual to the toric diagram, or, 
completely equivalently, is the projection of the 
polyhedral cone $\mathcal{C}$ onto the plane 
with normal vector $(1,0,0)$. The external legs of the  $(p,q)$ web are 
easily computed to be 
 \beq\label{legs} 
\begin{array}{l} 
(p_1,q_1)=(-c,-al) \\ 
(p_2,q_2)=(c-b,a(k+l)) \\ 
(p_3,q_3)=(b,-ak+1) \\ 
(p_4,q_4)=(0,-1) 
\end{array}~. 
\eeq 
This (p,q)-web is pictured in \fref{pq_generic}.
Using this information we can compute the total number of fields in 
the gauge theory. Specifically we have 
\beq 
N_{f}={1\over 2}\sum_{i,j\in \mbox{legs}}^4 \left| \det \left( \begin{array}{cc} p_i & q_i \\ 
p_j & q_j \end{array}\right) \right|~. 
\eeq 
This formula comes from computing intersection numbers of 
3--cycles in the mirror geometry \cite{Hanany:2001py}. 
In fact the four adjacent legs each contribute $a,b,c,d$ fields, 
which are simply the GLSM charges, up to sign. The two cross 
terms then contribute $c-a$ and $b-c$ fields, giving 
\bea 
N_{f} = a+3b~. 
\eea 
To summarise this section so far, the gauge theory for 
$L^{a,b,c}$ has $N_g=a+b$ gauge groups, and $N_f=a+3b$ fields in 
total. In Section 3.2 we will determine the 
multiplicities of the fields, as well as their baryon and 
flavour charges, using the results of the previous section. 
  
\subsection{The sub--family $L^{a,b,a}$} 
  
\label{subsection_Laba_geometry} 
  
The observant reader will have noticed that the charges in 
(\ref{legs}) are not always primitive. In fact this is a 
consequence of orbifold singularities in the Sasaki--Einstein space. 
In such singular cases one can have some number 
of lattice points, say $m-1$, on the edges of the toric diagram, and then 
the corresponding leg of the $(p,q)$ web in (\ref{legs}) is not 
a primitive vector. One should then really write the primitive 
vector, and associate to that leg the label, or multiplicity, $m$. 
Each leg of the $(p,q)$ web corresponds to a circle on $Y$ which is a locus 
of singular points if $m>1$, where $m$ gives the order of the 
orbifold group. Nevertheless, the charges (\ref{legs}) as written 
above give the correct numbers of fields. 
In fact this 
discussion is rather similar to the classification of 
\emph{compact} toric orbifolds in \cite{LT}, where each 
facet is assigned a positive integer label that describes the 
order of an orbifold group. Moreover, the non--primitive 
vectors are then used in the symplectic quotient. 
  
\begin{figure}[ht] 
  \epsfxsize = 11cm 
  \centerline{\epsfbox{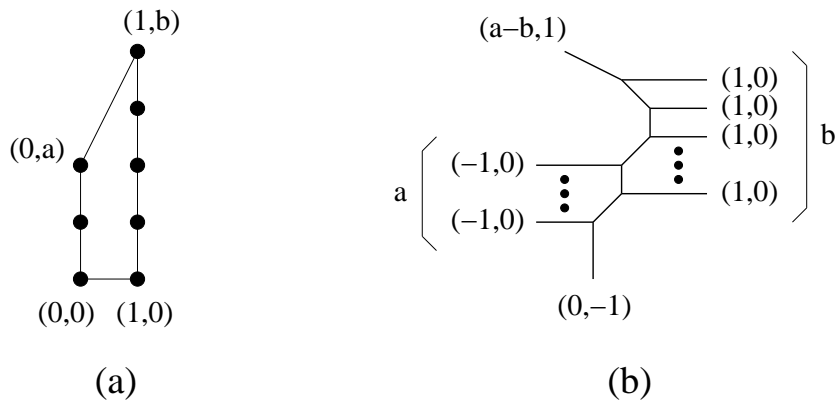}} 
  \caption{a) Toric diagram and b) $(p,q)$ web for the $L^{a,b,a}$ sub--family.} 
  \label{toric_web_Laba} 
\end{figure}

Rather than explain this point in generality, it is easier to 
give an example. Here we consider the family $L^{a,b,a}$, which 
are always singular if one of $a$ or $b$ is greater than 1. 
In fact by the $SL(2;\IZ)$ transformation 
\bea 
\left(\begin{array}{cc} 1 & l\\ 0& 1\end{array}\right)\eea 
one maps the toric diagram to an isosceles trapezoid as shown in \fref{toric_web_Laba}.a. 
  
Notice that there are $a-1$ lattice points on one external edge, 
and $b-1$ lattice points on the opposite edge. This is 
indicative of the singular nature of these spaces. 
Correspondingly, the $(p,q)$ web has non--primitive charges (or else 
one can assign positive labels $b$ and $a$ to primitive charges). 
Indeed, the leg with label $a$ is just the submanifold obtained by 
setting $Z_3=Z_4=0$. On $Y$ the D--terms, modulo the $U(1)$ gauge 
transformation, just give a circle $S^1$. However, the $U(1)$ 
group factors through $a$ times, due to the charges of $Z_1$ and $Z_2$ 
being both equal to $a$. This means that the $S^1$ is a locus 
of $\IZ_a$ orbifold singularities. Obviously, similar remarks apply 
to $Z_1=Z_2=0$. The singular nature of these spaces will 
also show up in the gauge theory: certain types of fields will be 
absent, and there will be adjoints, as well as bifundamentals. 
The $L^{a,b,a}$ family will be revisited in Section \ref{section_Laba_tilings}, 
where we will construct their associated brane tilings, gauge theories and compare 
the computations performed in the field theories with those in the dual supergravity 
backgrounds. 
  
\subsection{Quantum numbers of fields} 
  
Let us denote the distinguished fields as 
\bea 
X_1 = Y \qquad X_2= U_1 \qquad X_3 = Z \qquad X_4 = U_2~. 
\eea 
In the limit $c=d=p$ we have that $L^{a,b,c}$ reduces to a $Y^{p,q}$. 
Specifically, $b=p+q$ and $a=p-q$. Then this 
notation for the fields coincides with that of reference \cite{Benvenuti:2004dy}. 
In particular, the $U_i$ become a doublet under the $SU(2)$ 
isometry/flavour symmetry in this limit. 
  
The multiplicities of the fields can be read off from the 
results of the last section: 
\bea 
\mathrm{mult} [Y] = b  \qquad \mathrm{mult} [U_1] = d \qquad 
\mathrm{mult} [Z] = a \qquad \mathrm{mult} [U_2] = c~. 
\eea 
This accounts for $2(a+b)$ fields, which means that 
there are $b-a$ fields missing. The $(p,q)$ web suggests that there 
are two more fields $V_1$ and $V_2$ with multiplicities 
\bea 
\mathrm{mult} [V_1] = c-a\qquad 
\mathrm{mult} [V_2] = b-c~. 
\eea 
Indeed, this also reproduces a $Y^{p,q}$ theory in the 
limit $c=d$, where the fields $V_i$ again become an 
$SU(2)$ doublet. 
  
It is now simple to work out which toric divisors these 
additional fields are associated to. As will be explained 
later, each divisor must appear precisely $b$ times in the 
list of fields. Roughly, this is because there are necessarily 
$a+3b-(a+b)=2b$ terms in the superpotential, and 
every field must appear precisely twice by the quiver toric condition \cite{Feng:2002zw}. 
From this we deduce that we may view the remaining fields $V_1,V_2$ as ``composites'' --- 
more precisely, we identify them with unions 
of adjacent toric divisors $D_i \cup D_j$ in the Calabi--Yau, 
or equivalently in terms of supersymmetric 3--submanifolds in the Sasaki--Einstein space: 
\bea 
V_1: & & \qquad \Sigma_3 \cup \Sigma_4 \nn\\ 
V_2: & & \qquad \Sigma_2 \cup \Sigma_3~. 
\label{Vfields} 
\eea 

We may now compute the baryon and flavour charges of all the fields. 
The charges for the 
fields $V_1,V_2$ can be read off from their relation to the 
divisors $\Sigma_A$ above. We summarise the various quantum numbers 
in Table \ref{chargesgen}. 
\begin{table}[!h] 
\begin{center} 
$$\begin{array}{|c|c|c|c|c|c|}  \hline 
\mathrm{Field}& \mathrm{SUSY\ submanifold} &\mathrm{number}& U(1)_B &  U(1)_{F_1} & U(1)_{F_2} \\ \hline\hline 
Y & \Sigma_1 & b & a & 1 & 0 \\ \hline 
U_1   & \Sigma_2 & d & -c & 0& l \\ \hline 
Z & \Sigma_3 & a &  b & 0 & k \\ \hline 
U_2   & \Sigma_4 & c & -d & -1 & -k-l \\ \hline 
V_1 & \Sigma_3\cup\Sigma_4 & b-c & c-a & -1 & -l \\ \hline 
V_2 & \Sigma_2\cup\Sigma_3 & c-a & b-c & 0 & k+l \\ \hline 
\end{array}$$ 
\caption{Charge assignments for the six different types of fields present 
 in the general quiver diagram for $L^{a,b,c}$.} 
\label{chargesgen} 
\end{center} 
\end{table} 
  
Notice that the $SL(2;\IZ)$ transformation (\ref{icecream}) 
that shifts $k$ and $l$ is 
equivalent to redefining the flavour symmetry 
\bea 
U(1)_{F_2}\rightarrow U(1)_{F_2}-tU(1)_B+ta U(1)_{F_1}~.\eea 
Note also that each toric divisor appears precisely $b$ times in the table. This fact 
automatically ensures that the linear traces vanish 
\bea 
\mathrm{Tr} U(1)_B = 0\quad\qquad 
\mathrm{and} 
\qquad\quad 
\mathrm{Tr} U(1)_{F_1} = \mathrm{Tr} U(1)_{F_2} =0~, 
\eea 
as must be the case. As a non--trivial check of these assignments, 
one can compute that the cubic baryonic trace vanishes as well 
\bea 
\mathrm{Tr} U(1)_B^3 =  b a^3 -d c^3 + a b^3 - c d^3  + (b-c) (c-a)^3 + 
(c-a) (b-c)^3 = 0~. 
\eea

\subsection{The geometry} 
  
\label{section_metrics} 
  
In this subsection we summarise some aspects of the geometry of the 
toric Sasaki--Einstein manifolds $L^{a,b,c}$. First, we recall the 
metrics \cite{Cvetic:2005ft}, and how these are associated to the toric 
singularities discussed earlier \cite{Martelli:2005wy}. We also discuss 
supersymmetric submanifolds, compute their volumes, and use these 
results to extract the R--charges of the dual field theory. 
  
The local metrics were given in \cite{Cvetic:2005ft} in the form 
\bea\label{popes} \dd s^2 & = & \frac{\rho^2\dd 
x^2}{4\Delta_x}+\frac{\rho^2\dd\theta^2} {\Delta_{\theta}} + 
\frac{\Delta_x}{\rho^2}\left(\frac{\sin^2\theta}{\alpha} 
\dd\phi+\frac{\cos^2\theta}{\beta}\dd\psi\right)^2\nn\\ 
&& 
+ \frac{\Delta_{\theta}\sin^2\theta\cos^2\theta}{\rho^2}\left( 
\frac{\alpha-x}{\alpha}\dd \phi-\frac{\beta-x}{\beta}\dd\psi\right)^2+ (\dd \tau 
+ \sigma)^2 
\eea 
where 
\bea 
\sigma & = & \frac{\alpha-x}{\alpha}\sin^2\theta \dd \phi + 
\frac{\beta-x}{\beta}\cos^2\theta \dd \psi\nn\\ 
\Delta_x & = & x(\alpha-x)(\beta-x)-\mu,\qquad \rho^2=\Delta_{\theta}-x\nn\\ 
\Delta_{\theta} & = & \alpha\cos^2\theta +\beta\sin^2\theta~.\eea 
Here $\alpha,\beta,\mu$ are \emph{a priori} arbitrary constants. 
These local metrics are Sasaki--Einstein which can be equivalently stated 
by saying that the metric cone 
$\dd r^2 + r^2 \dd s^2$ is Ricci--flat and K\"ahler, or that the 
four--dimensional part of the metric (suppressing the $\tau$ direction) is 
a local K\"ahler--Einstein metric of positive curvature. These local 
metrics were also found in \cite{Martelli:2005wy}. 
%
The coordinates in (\ref{popes}) have the following ranges: $0\leq \theta \leq 
\pi/2$, $0 \leq \phi \leq 2\pi$, $0 \leq \psi \leq 2\pi$, and $x_1 \leq x \leq 
x_2$, where $x_1, x_2$ are the smallest two roots of the cubic 
polynomial $\Delta_x$. The coordinate $\tau$, which parameterises 
the orbits of the Reeb Killing 
vector $\de / \de \tau$ is generically non--periodic. In particular, generically the orbits 
of the Reeb vector field do not close, 
implying that the Sasaki--Einstein manifolds 
are in general \emph{irregular}. 
  
The metrics are clearly toric, meaning that there is a $U(1)^3$ 
contained in the 
isometry group. Three 
commuting Killing vectors are simply given by $\de/\de \psi, \de / \de \psi,\de 
/\de \tau$. The global properties of the spaces are then conveniently described 
in terms of those linear combinations of the vector fields that vanish over real codimension two 
fixed point sets. This will correspond to toric divisors in the Calabi--Yau 
cone --- see 
\emph{e.g.} \cite{Martelli:2004wu}. It is shown in \cite{Cvetic:2005ft} that there are precisely 
\emph{four} such vector fields, and in particular these are $\de / \de \phi$ 
and $\de / \de \psi$, vanishing on $\theta=0$ and $\theta=\pi/2$ respectively, 
and two additional vectors 
\bea 
\ell_i = a_i \frac{\de}{\de \phi} + b_i \frac{\de}{\de \psi} + c_i 
\frac{\de}{\de \tau}  \quad\qquad i=1,2 
\eea 
which vanish over $x=x_1$ and $x=x_2$, respectively. The constants are 
given by \cite{Cvetic:2005ft} 
\bea 
a_i = \frac{\alpha c_i}{x_i-\alpha}~, \qquad \qquad 
b_i = \frac{\beta c_i}{x_i-\beta}~,\nn\\ 
c_i = \frac{(\alpha-x_i)(\beta - 
x_i)}{2(\alpha+\beta)x_i-\alpha\beta-3x_i^2}~.\qquad 
\label{relations} 
\eea 
In order that the corresponding space is globally well--defined, there must be a 
linear relation 
between the four Killing vector fields 
\bea 
a \, \ell_1 + b \,\ell_2 + c \,\frac{\de}{\de \phi} +  d\,\frac{\de}{\de \psi} = 0 
\label{linrel} 
\eea 
where $(a,b,c,d)$  are relatively prime integers. 
It is shown in \cite{Cvetic:2005ft} that for appropriately chosen coefficients 
$a_i,b_i,c_i$  there are then countably infinite families of complete 
Sasaki--Einstein manifolds. 
  
The fact that there are four Killing vector fields that vanish on 
codimension 2 submanifolds implies that the image of the Calabi--Yau cone 
under the moment map for the ${T}^3$ action is a four faceted polyhedral 
cone in $\IR^3$ \cite{Martelli:2004wu}. Using the linear relation among the 
vectors (\ref{linrel}) one can show that the normal vectors to this 
polyhedral cone  satisfy the relation 
\be 
a \, v_1-c\, v_2+b\,v_3-(a+b-c)\,v_4=0\ee 
where $v_A$, $A=1,2,3,4$ are the primitive vectors in $\IR^3$ that define 
the cone. Note that we have listed  the vectors according to the order of the 
facets of the polyhedral cone. As explained in \cite{Martelli:2005wy}, 
it follows that, for $a,b,c$ relatively prime, 
the Sasaki--Einstein manifolds arise from the symplectic quotient 
\be\label{symplectic} 
{\IC}^4//(a,-c,b,-a-b+c)~\ee 
which is precisely the gauged linear sigma model considered in the 
previous subsection.

The volume of the Sasaki--Einstein manifolds/orbifolds is given by 
\cite{Cvetic:2005ft} \bea \vol (Y)& = & \frac{\pi^2}{2k\alpha\beta} 
(x_2-x_1)(\alpha+\beta -x_1-x_2) \Delta\tau \eea where here 
$k=$gcd$(a,b)$ and \be \Delta \tau = \frac{2\pi k |c_1|}{b}~. \ee 
This can also be written as \bea \vol (Y)& = & 
\frac{\pi^3(a+b)^3}{8abcd}W \label{volume_quartic} \eea where $W$ is 
a root of certain quartic polynomial given in \cite{Cvetic:2005ft}. This 
shows that the central charges of the dual conformal field theory 
will be generically \emph{quartic irrational}. 
  
In order to compute the R--charges from the metric, we need to know 
the volumes of the four supersymmetric 3--submanifolds $\Sigma_A$. 
These volumes were not given in \cite{Cvetic:2005ft} but it is 
straightforward to compute them. We obtain \bea \vol (\Sigma_1) = 
\frac{\pi}{k} \left|\frac{c_1}{a_1b_1}\right| \Delta \tau\qquad 
\vol (\Sigma_2)  =  \frac{\pi}{k\beta}(x_2-x_1)\Delta \tau ~\;\nn\\ 
\vol (\Sigma_3)  =  \frac{\pi}{k} \left|\frac{c_2}{a_2b_2}\right| \Delta \tau\qquad 
\vol (\Sigma_4)  =  \frac{\pi}{k\alpha}(x_2-x_1)\Delta \tau ~. 
\label{susyvolumes} 
\eea 
We can now complete the charge assignments of all the fields in the 
quiver by giving their R--charges purely from the geometry. The 
charges of the distinguished fields $Y,U_1,Z,U_2$ are obtained from 
the geometry using the formula \bea R[X_A] & = & \frac{\pi}{3} 
\frac{\vol (\Sigma_A)}{\vol (Y)}~, \eea while those of the $V_1, 
V_2$ fields are simply deduced from (\ref{Vfields}). In particular 
\bea R[V_1] = R[Z]+ R[U_2]\qquad \qquad R[V_2]  =  R[Z]+ R[U_1]~. 
\eea 
  
It will be convenient to note that the constants $\alpha,\beta,\mu$ 
appearing in $\Delta_x$ are related to its roots as follows \bea 
\mu & = & x_1 x_2 x_3\nn\\ 
\alpha + \beta & = & x_1 + x_2 + x_3\nn\\ 
\alpha \beta & = & x_1 x_2 + x_1 x_3 + x_2 x_3 ~, \label{roots} \eea 
where $x_3$ is the third root of the cubic, and $x_3\geq x_2 \geq 
x_1 \geq 0$. Using the volumes in (\ref{susyvolumes}), we then 
obtain the following set of R--charges 
%
\bea 
R[Y] = \frac{2}{3x_3}(x_3-x_1)\qquad 
\qquad  R[U_1] = \frac{2\alpha}{3x_3}~\;\nn\\ 
R[Z] = \frac{2}{3x_3}(x_3-x_2)\qquad \qquad  R[U_2] = 
\frac{2\beta}{3x_3}~. \label{eqs_geometric_1} \eea To obtain 
explicit expressions, one  should now write the constants 
$x_i,\alpha,\beta$ in terms of the integers $a,b,c$. This can be 
done, using the equations (9) in \cite{Cvetic:2005ft}. We have 
\bea 
\frac{x_1 (x_3-x_1)}{x_2 (x_3-x_2)} & = & \frac{a}{b}\;\;\\ 
\frac{\alpha (x_3-\alpha)}{\beta (x_3-\beta)} & = & \frac{c}{d}~. 
\eea 
Notice that  $\alpha=\beta$ implies $c=d$, as claimed in 
\cite{Cvetic:2005ft}. Combining these two equations with (\ref{roots}), one 
obtains a complicated system of quartic polynomials, which in 
principle can be solved. However, we will proceed differently. Our 
aim is simply to show that the resulting R--charges will match with 
the $a$--maximisation computation in the field theory. Therefore, 
using the relations above, we can write down a system of equations 
involving the R--charges and the integers $a,b,c,d$. We obtain the 
following: 
 \bea 
\frac{R[Y](2-3 R[Y])}{R[Z](2-3 R[Z])} & = & \frac{a}{b} \nonumber 
\\[2mm] 
\frac{R[U_1](2-3 R[U_1])}{R[U_2](2-3 R[U_2])} & = & \frac{c}{d} \nonumber \\ 
\frac{3}{4}\left(R[U_1]R[U_2] -R[Y]R[Z]\right) + R[Z] +R[Y] & = & 1 \nonumber \\ 
R[Y]+ R[U_1]+ R[Z] + R[U_2] &  = & 2~. \label{rcharges} \eea With 
the aid of a computer program, one can check that the solutions to 
this system are given in terms of roots of various quartic 
polynomials involving $a,b,c$. For the case of $L^{a,b,a}$ the 
polynomials reduce to quadratics and the R--charges can be given in 
closed form. These in fact match precisely with the values that we 
will compute later using $a$--maximisation, as well as 
$Z$--minimisation. Therefore we won't record them here. 
  
In the general case, instead of giving the charges in terms of 
unwieldy quartic roots, we can more elegantly show that the system 
(\ref{rcharges}) can be recast into an equivalent form which is 
obtained from $a$--maximisation. In order to do so, we can use the 
last equation to solve for $R[U_1]$. Expressing the first three 
equations in terms of $R[U_2]=x$, $R[Y]=y$ and $R[z]=z$, we have 
\beq 
\begin{array}{rcl} 
b(2-3y)+az(3z-2)&=&0 \\ 
c(x+y-2)(3x+3y-4)-(a+b-c)(x-z)(3x-3z-2)&=&0 \\ 
3x^2-4y+2(z+2)+x(3y-3z-6)&=&0\, . 
\end{array} 
\label{eqs_geometric_2} 
\eeq 
Interestingly, the third equation does not involve any of the 
parameters. For later comparison with the results coming from 
$a$--maximisation, it is important to find a way to reduce this 
system of three coupled quadratic equations in three variables to a 
standard form. The simplest way of doing so is to `solve' for one of 
the variables $x$, $y$ or $z$ and two of the parameters. A 
particularly simple choice is to solve for $y$, $a$ and $b$. The 
simplicity follows from the fact that it is possible to use the 
third equation to solve for $y$  and the parameters then appear linearly 
in all the equations. Doing this, we obtain 
\bea 
a&=&{c(3x-2)\left(3x(x-z-2)+2(2+z) \right) \over (3x-4)^2 (x-z)}  \nonumber \\ 
b&=&{c z (3z-2)\over (3x-3z-2)(x-z)} \nonumber \\ 
y&=&{-2 (2+z)-3x (x-z-2)\over (3x-4)}. \label{eqs_geometric_3} \eea 
This system of equations is equivalent to the original one, and is 
the one we will compare with the results of $a$--maximisation. 
  
Of course, one could also compute these R--charges using 
$Z$--minimisation \cite{Martelli:2005tp}. The algebra encountered in 
tackling the minimisation problem is rather involved, but it is straightforward 
to check agreement of explicit results on a case by case basis. 
  

\section{Superpotential and gauge groups} 
 
\label{section_superpotential_and_gauge} 
 
In the previous sections we have already described how rather 
generally one can obtain the number of gauge groups, and the field 
content of a quiver whose vacuum moduli space should reproduce the 
given toric variety. In particular, we have listed the 
multiplicities of every field and their complete charge assignments, 
namely their baryonic, flavour, and R--charges. In the following we 
go further and predict the form of the superpotential as well as the 
nature of the gauge groups, that is, the types of nodes appearing 
in the quivers.

\subsection{The superpotential} 
 
\label{subsection_superpotential} 
 
First, we recall that in \cite{Franco:2005rj} a general formula was derived 
relating the number of gauge groups $N_g$, the number of fields 
$N_f$, and the number of terms in the superpotential $N_W$. This 
follows from applying Euler's formula to a brane tiling that lives 
on the surface of a 2--torus, and reads \bea N_W & = & N_f - N_g\, . 
\eea Using this we find that the number of superpotential terms for 
$L^{a,b,c}$ is $N_W=2b$. Now we use the fact each term in the 
superpotential $W$ must be \bea\label{summy} \cup_{A=1}^4 \Sigma_A~. 
\eea In fact this is just the canonical class of $X$ -- a standard 
result in toric geometry. One can justify the above form as follows. 
Each term in $W$ is a product of fields, and each field is 
associated to a union of toric divisors. The superpotential has  
R--charge 2, and is uncharged under the baryonic and flavour 
symmetries. This is true, using the results of Section 2 and 
(\ref{summy}).

A quick inspection of Table \ref{chargesgen} then allows us to identify three types 
of monomials that may appear in the superpotential 
\bea 
W_{q} = \mathrm{Tr}\, YU_1ZU_2 \qquad W_{c_1} = \mathrm{Tr}\, YU_1V_1 
\qquad W_{c_2} = \mathrm{Tr}\, YU_2V_2~. 
\label{WBB} 
\eea 
Furthermore, their number is uniquely fixed by the mutiplicities of the fields, and the 
fact that $N_W=2b$. The schematic form of the superpotential for a general $L^{a,b,c}$ 
quiver theory is then 
\bea 
W & = & 2 \left[ a \, W_{q} + (b-c)\, W_{c_1} + (c-a)\,  W_{c_2}\right]~. 
\eea 
In the language of dimer models, this is telling us the types of vertices in the brane 
tilings \cite{Franco:2005rj}. In particular, in each fundamental domain of the tiling we must have 
$2a$ four--valent vertices, $2(b-c)$ three--valent 
vertices of type 1, and $2(c-a)$ three--valent vertices of type 2.

\subsection{The gauge groups} 
 
\label{subsection_nodes} 
 
Finally, we discuss the nature of the $N_g=a+b$ gauge groups of the gauge 
theory, i.e. we determine the types of nodes in the quiver. This information, together 
with the above, will be used to construct the brane tilings. First, we will identify 
the allowed types of nodes, and then we will determine the number of times each node 
appears in the quiver.

The allowed types of nodes can be deduced by requiring that at any given node 
\begin{enumerate} 
 
\item 
the total baryonic and flavour charge is zero: $\quad \sum_{i\in \mathrm{node}} U(1)_i=0$ 
 
\item 
the beta function vanishes: $\quad \sum_{i\in \mathrm{node}} (R_i -1) + 2 =0 $ 
 
\item 
there are an even number of legs. 
 
\end{enumerate} 
 
These requirements are physically rather obvious. The first property  is satisfied 
if we construct a node out of products (and powers) of the building blocks 
of the superpotential (\ref{WBB}). Moreover, using (\ref{W=2}), this also guarantees that 
the total R--charge at the node is even. 

Imposing these three requirements turns out to be rather 
restrictive, and we obtain \emph{four} different types of nodes that 
we list below: \bea A:\, U_1YV_1\cdot  U_1YV_1\quad B:\, 
V_2YV_1\cdot  U_1YU_2\quad D:\,  U_2YV_2\cdot  U_2YV_2\, \quad C:\, 
U_1YU_2Z \label{types_nodes} \eea 
 
Next, we determine the number of times each node appears in the 
quiver. Denote these numbers $n_A, n_B, n_D, 2n_C$ respectively. 
Taking into account the multiplicities of all the fields imposes six 
linear relations. However, it turns out that these \emph{do not} uniquely fix the number of different nodes. We have \bea 
n_C & = & a\nn\\ 
n_B + 2n_A & = & 2 (b-c)\nn\\ 
n_B + 2n_D & = & 2 (c-a)~. \label{AD} \eea 
 
Although the number of fields and schematic form of superpotential 
terms are fixed by the geometry, the number of $A$, $B$ and $D$ 
nodes are not. We can then have different types of quivers that are 
nevertheless described by the same toric singularity. This suggests 
that the theories with different types of nodes are related by 
Seiberg dualities. We will show that this is the case in Section 6.
 
It is interesting to see what happens for the $L^{a,b,a}$ geometries 
discussed in Section \ref{subsection_Laba_geometry}. In this case 
$c-a=0$  (the case that $b-c=0$ is symmetric with this) and the 
theory has some peculiar properties. Recall that this corresponds to 
a linear sigma model with charges $(a,-a,b,-b)$. These theories have 
no $V_2$ fields, while the $b-a$ $V_1$ fields have zero baryonic 
charge, and must therefore be \emph{adjoints}. Moreover, from 
(\ref{AD}), we see that $n_B=n_D=0$, so that there aren't any $B$ 
and $D$ type nodes, while there are $b-a$ $A$--type nodes. In terms 
of tilings, these theories are then just constructed out of 
$C$--type quadrilaterals and $A$--type hexagons. We will consider in 
detail these models in Section \ref{section_Laba_tilings}. 
 
Finally, we note that the general conclusions derived for 
$L^{a,b,c}$ quivers in Sections \ref{subsection_superpotential} and 
\ref{subsection_nodes} are based on the underlying assumption that 
we are dealing with a generic theory ({\it i.e.} one in which the 
R--charges of different types of fields are not degenerate). It is 
always possible to find at least one generic phase for a given 
$L^{a,b,c}$, and thus the results discussed so far apply. 
Non--generic phases can be generated by Seiberg duality 
transformations. In these cases, new types of superpotential 
interactions and quiver nodes may emerge, as well as new types of 
fields. This was for instance the case for the toric phases of the 
$Y^{p,q}$ theories \cite{Benvenuti:2004wx}.

\section{R--charges from  $a$--maximisation} 
 
\label{section_amax} 
 
A remarkable check of the AdS/CFT correspondence consists of 
matching the gauge theory computation of R--charges and central 
charge with the corresponding calculations of volumes of the 
dual Sasaki--Einstein manifold and supersymmetric submanifolds 
on the gravity side. This is perhaps the most convincing evidence 
that the dual field theory is the correct one. Since explicit 
expressions for the Sasaki--Einstein metrics are available, it is 
natural to attempt such a check. Actually, the volumes of toric 
manifolds and supersymmetric submanifolds thereof can also be 
computed from the toric data \cite{Martelli:2005tp}, without using a metric. 
This gives a third independent check that everything is indeed 
consistent.

Here we will calculate the R--charges and central charge $a$ for an 
arbitrary $L^{a,b,c}$ quiver gauge theory using a--maximisation. 
From the field theory point of view, initially, there are six 
different R--charges, corresponding to the six types of bifundamental 
fields $U_1$, $U_2$, $V_1$, $V_2$, $Y$ and $Z$. Since the field 
theories are superconformal, these R--charges are such that the beta 
functions for the gauge and superpotential couplings vanish. Using 
the constraints \eref{WBB} and \eref{types_nodes} it is possible to 
see that these conditions always leave us with a three--dimensional 
space of possible R--charges. This is in precise agreement with the 
fact that the non--R abelian global symmetry is $U(1)^3\simeq 
U(1)_F^2\times U(1)_B$. It is convenient to adopt the 
parametrization of R--charges of Section \ref{section_metrics}: 
\beq 
\begin{array}{lclclcl} 
R[U_1] & = & x-z  & \ \ \ \ \ & 
R[U_2] & = & 2-x-y \\ 
R[V_1] & = & 2-x-y+z & \ \ \ \ \ & 
R[V_2] & = & x \\ 
R[Y]   & = & y & \ \ \ \ \ & R[Z]   & = & z\, . 
\end{array} 
\label{R-parametrization} \eeq This guarantees that all beta 
functions vanish. Using the multiplicities in Table 
\ref{chargesgen}, we can check that $\tr R(x,y,z)=0$. This is 
expected, since this trace is proportional to the sum of all the 
beta functions. In addition, the trial $a$ central charge can be 
written as \bea \tr R^3(x,y,z)&=&{1\over3} 
\left[a\left(9(2-x)(x-z)z-2 
\right)+b \left(9\ x \ y \ (2-x-y)-2 \right)\right.\nonumber\\ 
&+& \left. 9(b-c) \ y \ z \ (2x+y-z-2) \right]. \label{a_general} 
\eea 
The R--charges are determined by maximising \eref{a_general} 
with respect to $x$, $y$ and $z$. 
This corresponds to the following equations 
\beq 
\begin{array}{lclcl} 
\partial_{x} \ \tr R^3(x,y,z) & = & 0 & = &-3 \ b \ y \ (2x+y-2z-2)+3 \ z \ \left( a(2-2x+z-2cy) \right) \\ 
\partial_{y} \ \tr R^3(x,y,z) & = & 0 & = & -3\ b \ (x-z)(x+2y-z-2)-3 \ c \ z (2x+2y-z-2) \\ 
\partial_{z} \ \tr R^3(x,y,z) & = & 0 & = & -3\ a \ (x-2)(x-2z)+3(b-c) \ y \ (2x+y-2z-2) 
\end{array} 
\label{eqs_amax_1}~. 
\eeq 
It is straightforward to show that this system of equations is 
equivalent to \eref{eqs_geometric_3}. In fact, proceeding as in 
Section \ref{section_metrics}, we reduce \eref{eqs_amax_1} to an 
equivalent system by `solving' for $y$, $a$ and $b$. In this case, 
there are three solutions, although only one of them does not 
produce zero R--charges for some of the fields, and indeed 
corresponds to the local maximum of \eref{a_general}. This solution 
corresponds to the following system of equations 
\bea 
a&=&{c(3x-2)\left(3x(x-z-2)+2(2+z) \right) \over (3x-4)^2 (x-z)}  \nonumber \\ 
b&=&{c z (3z-2)\over (3x-3z-2)(x-z)} \nonumber \\ 
y&=&{-2 (2+z)-3x (x-z-2)\over (3x-4)} \eea which is identical 
to \eref{eqs_geometric_3}. We conclude that, for the entire 
$L^{a,b,c}$ family, the gauge theory computation of R--charges and 
central charge using a--maximisation agrees precisely with the values 
determined using geometric methods on the gravity side of the 
AdS/CFT correspondence.

\section{Constructing the gauge theories using brane tilings} 
 
In Sections \ref{section_Labc} and \ref{section_superpotential_and_gauge} we derived 
detailed information regarding the gauge theory 
on D3--branes transverse to the cone over an arbitrary $L^{a,b,c}$ space. 
Table \ref{chargesgen} gives the types of fields along with their multiplicities and global $U(1)$ charges, 
\eref{WBB} presents the possible superpotential interactions and \eref{types_nodes} and \eref{AD} 
give the types of nodes in the quiver along with some constraints on their multiplicities. 
 
This information is sufficient for constructing the corresponding gauge theories. 
We have used it in Section \ref{section_amax} to prove perfect agreement between the geometric and gauge
theory
computations of R--charges and central charges.
Nevertheless, it is usually a formidable task to combine all these pieces of information to generate 
the gauge theory. In this section we introduce a simple set of rules for the construction 
of the gauge theories for the $L^{a,b,c}$ geometries. In particular, our goal is to find a simple procedure 
in the spirit of the `impurity idea' of \cite{Benvenuti:2004dy,Benvenuti:2004wx}. 
 
Our approach uses the concept of a {\bf brane tiling}, which was introduced in 
\cite{Franco:2005rj}, following the discovery of the connection between toric geometry and 
dimer models of \cite{Hanany:2005ve}. Brane tilings encode both the quiver diagram and the superpotential 
of gauge theories on D--branes probing toric singularities. Because of this simplicity, 
they provide the most suitable 
language for describing complicated gauge theories associated with toric geometries. We refer the reader 
to \cite{Franco:2005rj} for a detailed explanation of brane tilings and their relation to dimer models. 
 
\begin{figure}[ht] 
  \epsfxsize = 13cm 
  \centerline{\epsfbox{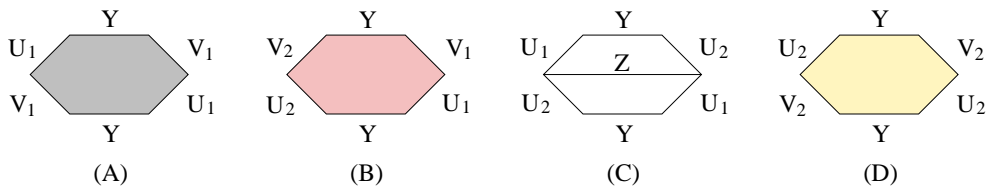}} 
  \caption{The four building blocks for the construction of brane tilings for $L^{a,b,c}$.} 
  \label{building_blocks} 
\end{figure}

All the conditions of Sections \ref{section_Labc} and \ref{section_superpotential_and_gauge} 
can be encoded in the properties of four elementary building blocks. These blocks are shown 
in \fref{building_blocks}. We denote them A, B, C and D, following the corresponding labeling 
of gauge groups in \eref{types_nodes}. It is important to note that a C hexagon contains two nodes of 
type C. 
 
Every edge in the elementary hexagons is associated with a particular type of field. These edge labels 
fully determine the way in which hexagons can be glued together along their edges to form a periodic tiling. The quiver 
diagram and superpotential can then be read off from the resulting tiling using the results of \cite{Franco:2005rj}. 
The elementary hexagons automatically incorporate the three 
superpotential interactions of \eref{WBB}. The number of A, B, C and 
D hexagons is $n_A$, $n_B$, $n_C$ and $n_D$, respectively. Taking 
their values as given by \eref{AD}, the multiplicities in Table 
\ref{chargesgen} are reproduced. 
 
Following the discussion in Section 
\ref{section_superpotential_and_gauge}, the 
number of A, B, C and D hexagons is not fixed for a given $L^{a,b,c}$ geometry.
There is a one 
parameter space of solutions to \eref{AD}, which we can take to be 
indexed by $n_D$. It is possible to go from one solution of 
\eref{AD} to another one by decreasing the number of B hexagons by two and 
introducing one A and one D, {\it i.e.} $(n_A,n_B,n_C,n_D)\rightarrow 
(n_A+1,n_B-2,n_C,n_D+1)$. We show in the next 
section that this freedom in the number of each type of hexagon is 
associated with Seiberg duality.

\subsection{Seiberg duality and transformations of the tiling} 
 
We now study Seiberg duality \cite{Seiberg:1994pq} transformations that produce `toric quivers' 
\footnote{This term was introduced in \cite{Feng:2000mi} and refers to quivers in which the ranks 
of all the gauge groups are equal. Toric quivers are a subset of the infinite set of 
Seiberg dual theories associated to a given toric singularity, {\it i.e.} it is possible to obtain quivers 
that are not toric on D--branes probing toric singularities.}. 
We can go from one toric quiver to another one by applying Seiberg duality 
to the so--called {\bf self--dual} nodes. These are nodes for which the number 
of flavours is twice the number of colours, thus ensuring
that the rank of the dual gauge group does not change after
Seiberg duality. Such nodes are represented by squares in the brane tiling \cite{Franco:2005rj}. 
Hence, for $L^{a,b,c}$ theories, we only have to consider dualizing C nodes. 
Seiberg duality on a self--dual node corresponds to a {\bf local} transformation 
of the brane tiling \cite{Franco:2005rj}. This is important, since it means that 
we can focus on the sub--tilings surrounding the nodes of interest in order to 
analyse the possible behavior of the tiling. 
  
We will focus on cases in which the tiling that results from dualizing a self--dual node 
can also be described in terms of A, B, C and D hexagons. 
There are some cases in which 
Seiberg duality generates tilings that are not constructed using the elementary 
building blocks of \fref{building_blocks}. In these cases, the assumption of 
the six types of fields being non--degenerate does not hold. We present an example 
of this non-generic case
below, corresponding to $L^{2,6,3}$. 
 
\begin{figure}[ht] 
  \epsfxsize = 13cm 
  \centerline{\epsfbox{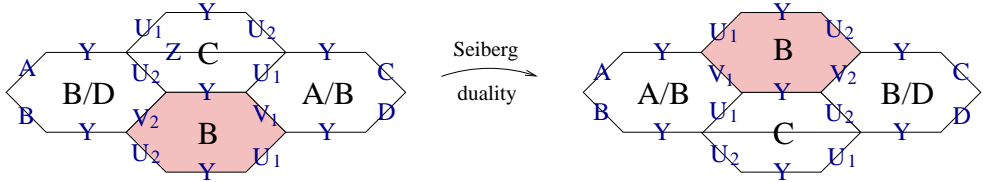}} 
  \caption{Seiberg duality on a self--dual node that does not change the hexagon content.} 
  \label{seiberg1} 
\end{figure}

\begin{figure}[ht] 
  \epsfxsize = 13cm 
  \centerline{\epsfbox{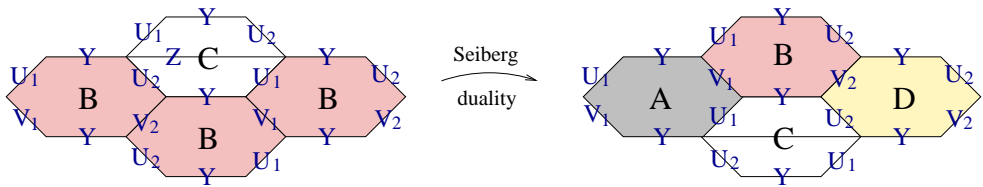}} 
  \caption{Seiberg duality on a self--dual node under which $(n_A,n_B,n_C,n_D) 
  \rightarrow(n_A+1,n_B-2,n_C,n_D+1)$.} 
  \label{seiberg2} 
\end{figure}

Leaving aside non--generic cases, we see that we only need to consider 
two possibilities. They are presented in Figures \ref{seiberg1} and \ref{seiberg2}.
\fref{seiberg1} shows the case in which dualization of the central 
square does not change the number of hexagons of each type in the 
tiling. The new tiling is identical to the original one up to a 
shift. \fref{seiberg2} shows a situation in which dualization of the central square 
removes two type B hexagons and adds an A and a D. 
 
The operations discussed above leave the labels of the edges on the boundary of the sub--tiling invariant 
\footnote{There is a small subtlety in this argument: in some cases, identifications of faces due to the periodicity 
of the tiling can be such that the boundary of the sub--tiling is actually modified when performing Seiberg duality.}. 
Hence, the types of hexagons outside the sub--tilings are not modified. 
The above discussion answers the question of how to interpret the 
different solutions of \eref{AD}: they just describe Seiberg dual 
theories. 
 
\subsection{Explicit examples} 
 
Having presented the rules for constructing tilings for a given $L^{a,b,c}$, we 
now illustrate their application with several examples. We first consider $L^{2,6,3}$, which 
is interesting since it has eight gauge groups and involves A, B and C 
elementary hexagons. We also discuss how its tiling is transformed under the action of 
Seiberg duality. We then present tilings for $L^{2,6,4}$, showing how D hexagons are generated by
Seiberg duality. Finally, we classify all sub--families whose brane tilings can be constructed using 
only two types of elementary hexagons. These theories are particularly simple and 
it is straightforward to match the geometric and gauge theory computations of R--charges 
and central charges explicitly. We analyse the $L^{a,b,a}$ sub--family in detail,
and present other interesting examples in the appendix. 
 
\subsubsection{Gauge theory for $L^{2,6,3}$} 
 
\begin{figure}[ht] 
  \epsfxsize = 12cm 
  \centerline{\epsfbox{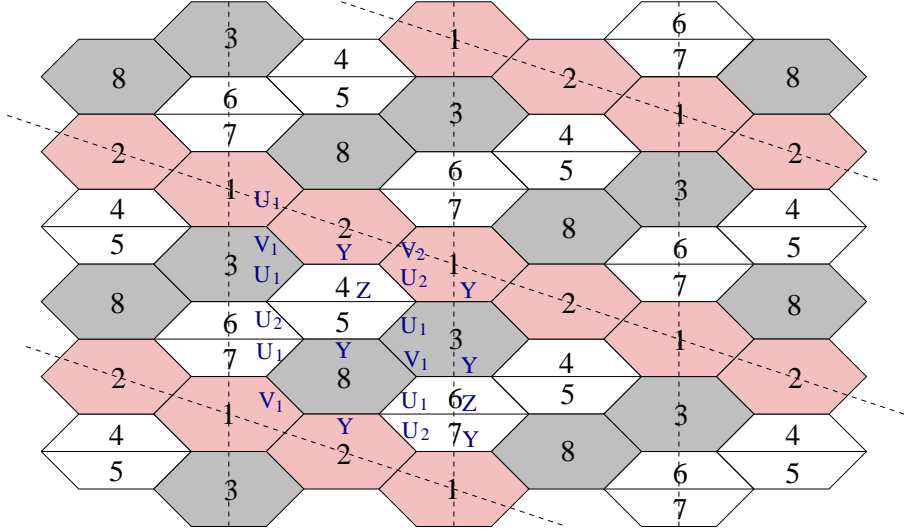}} 
  \caption{Brane tiling for $L^{2,6,3}$.} 
  \label{Y421} 
\end{figure}

Let us construct the brane tiling for $L^{2,6,3}$. We consider the 
$(n_A,n_B,n_C,n_D)=(2,2,2,0)$ solution to \eref{AD}. Hence, we have 
two A, two B and two C hexagons. 
Using the gluing rules given by the edge labeling in \fref{building_blocks}, it is straightforward 
to construct the brane tiling shown in \fref{Y421}. 
 
\begin{figure}[ht] 
  \epsfxsize = 7cm 
  \centerline{\epsfbox{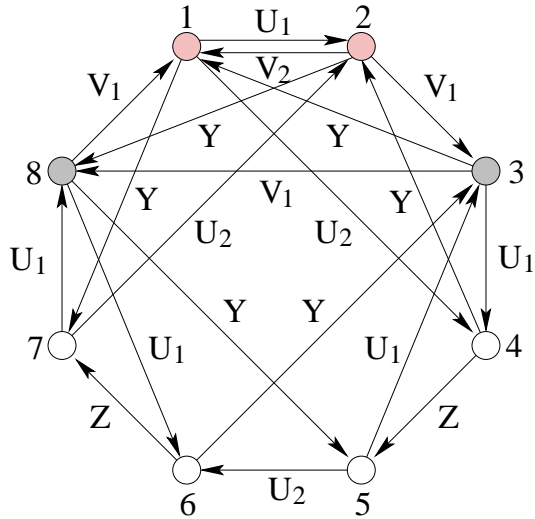}} 
  \caption{Quiver diagram for $L^{2,6,3}$.} 
  \label{quiver_Y421} 
\end{figure} 
 
From the tiling 
we determine the quiver diagram shown in \fref{quiver_Y421}. The multiplicities of each type 
of field are in agreement with the values in Table 2. 
In addition, we can also read off the corresponding superpotential 
 
\beq 
\begin{array}{rcl} 
W & = & Y_{31}U^{(1)}_{12}V^{(1)}_{23}-Y_{42}V^{(1)}_{23}U^{(1)}_{34}+Y_{42}V^{(2)}_{21}U^{(2)}_{12}+Y_{85}U^{(1)}_{53}V^{(1)}_{38} \\ 
  & + & Y_{17}U^{(1)}_{78}V^{(1)}_{81}-Y_{63}V^{(1)}_{38}U^{(1)}_{86}-Y_{28}V^{(1)}_{81}U^{(1)}_{12}-Y_{17}U^{(2)}_{72}V^{(2)}_{21} \\ 
  & + & Z_{45}U^{(2)}_{56}Y_{63}U^{(1)}_{34}-Z_{45}U^{(1)}_{53}Y_{31}U^{(2)}_{14}-Z_{67}U^{(1)}_{78}Y_{85}U^{(2)}_{56}+Z_{67}U^{(2)}_{72}Y_{28}U^{(1)}_{86} 
\end{array} 
\eeq 
where for simplicity we have indicated the type of $U$ and $V$ fields with a superscript and have used subscripts for the  
gauge groups under which the bifundamental fields are charged. 
 
\begin{figure}[ht] 
  \epsfxsize = 3cm 
  \centerline{\epsfbox{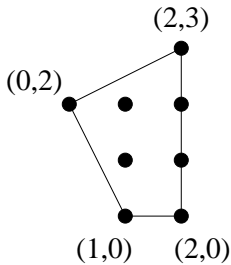}} 
  \caption{Toric diagram for $L^{2,6,3}$ determined using the characteristic polynomial of the Kasteleyn matrix for the tiling in \fref{Y421}.} 
  \label{toric_L263} 
\end{figure}

Having the brane tiling for a gauge theory at hand makes the derivation of its moduli space straightforward. The corresponding toric diagram  is determined from the characteristic polynomial of the Kasteleyn matrix of the tiling \cite{Hanany:2005ve,Franco:2005rj}. In this case, we obtain the toric diagram shown in \fref{toric_L263}. This is an additional check of our construction. 
 
\begin{figure}[ht] 
  \epsfxsize = 10cm 
  \centerline{\epsfbox{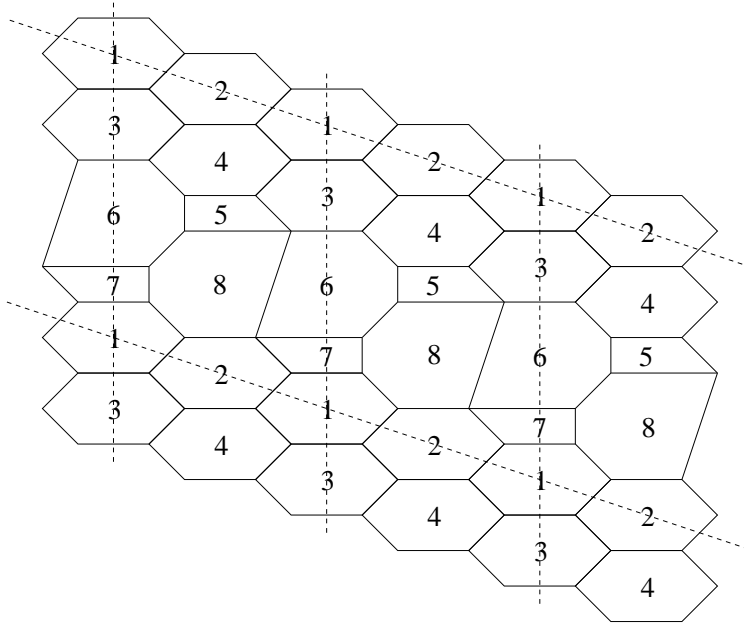}} 
  \caption{Brane tiling for a Seiberg dual phase of $L^{2,6,3}$.} 
  \label{L263_2} 
\end{figure}

Let us now consider how Seiberg duality on self--dual nodes acts on this tiling. Dualization of nodes 4 or 7 corresponds to the 
situation in \fref{seiberg1}. The resulting tiling is identical to the original one up to an upward or downward shift, respectively. 
The situation is different when we dualize node 5 or 6. In these cases, Seiberg duality `splits apart' the two squares corresponding 
to the C nodes forming C hexagons. \fref{L263_2} shows the tiling after dualizing node 5. 
This tiling seems to violate the classification of possible gauge groups given in Section \ref{WBB}. 
In particular, some of the hexagons would have at least one edge corresponding to a Z field. As we discussed in Section \ref{subsection_nodes}, 
this is not a contradiction, but just indicates that we are in a non--generic situation in which 
some of the six types of fields are degenerate. 
 
\subsubsection{Generating D hexagons by Seiberg duality: $L^{2,6,4}$} 

\begin{figure}[ht] 
  \epsfxsize = 7cm 
  \centerline{\epsfbox{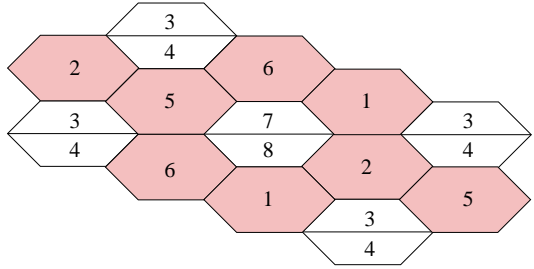}} 
  \caption{Brane tiling for $L^{2,6,4}$.} 
  \label{L264_1} 
\end{figure}

We now construct brane tilings for $L^{2,6,4}$. This geometry is actually a $\IZ_2$ orbifold of $L^{1,3,2}$.
This example illustrates how D hexagons are generated by Seiberg duality. We start with the
$(n_A,n_B,n_C,n_D)=(0,4,2,0)$ solution to \eref{AD}, whose corresponding tiling is shown in 
\fref{L264_1}.

\begin{figure}[ht] 
  \epsfxsize = 7cm 
  \centerline{\epsfbox{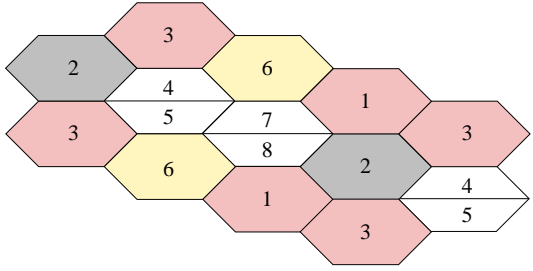}} 
  \caption{Brane tiling for $L^{2,6,4}$.} 
  \label{L264_2} 
\end{figure}

We see that all self-dual nodes are of the form presented in \fref{seiberg2}. Seiberg duality on node 4
leads to a tiling with $(n_A,n_B,n_C,n_D)=(1,2,2,1)$, which we show in \fref{L264_2}.

\subsubsection{The $L^{a,b,a}$ sub--family} 
 
\label{section_Laba_tilings} 
 
It is possible to use brane tilings to identify infinite 
sub--families of the $L^{a,b,c}$ theories whose study is considerably 
simpler than the generic case. In particular, classifying the 
geometries whose corresponding tilings can be constructed using only 
two different types of hexagons is straightforward. We now proceed 
with such a classification. 
 
Let us first consider those models that do not involve C type 
hexagons. These tilings consist entirely of `pure' hexagons and thus 
correspond to orbifolds \cite{Hanany:2005ve,Franco:2005rj}. We have 
already discussed them in Section \ref{section_Labc}, where we 
mentioned the case in which $a$, and thus $n_C$, is equal to zero. 
The orbifold action is determined by the choice of a 
fundamental cell (equivalently, by the choice of labeling of faces 
in the tiling). 
 
We now focus on theories for which one of the two types of hexagons is of type 
C. There are only three possibilities of this form: 
\beq 
\begin{array}{ccc} 
& \ \ \ \mbox{{\bf Hexagon types}} \ \ \  &  \ \ \ \mbox{{\bf Sub--family}} \ \ \  \\ 
n_B=n_D=0 & \mbox{A and C} & L^{a,b,a} \\ 
n_A=n_D=0 & \mbox{B and C} & L^{a,b,{a+b\over 2}}=Y^{{a+b\over 2},{a-b\over 2}} \\ 
n_A=n_B=0 & \mbox{C and D} & L^{a,b,b} 
\end{array} 
\eeq 
It is interesting to see that the $Y^{p,q}$ theories emerge naturally from 
this classification of simple models. In addition to orbifolds and $Y^{p,q}$'s, 
the only new family is that of $L^{a,b,a}$. The $L^{a,b,b}$ 
family is equivalent to the latter by a trivial reordering of the GLSM charges, which in the gauge theory 
exchanges $U_1 \leftrightarrow U_2$ and $V_1 \leftrightarrow V_2$. 
 
Let us study the gauge theories for the $L^{a,b,a}$ manifolds. These theories were first studied in \cite{Uranga:1998vf} using Type IIA configurations of relatively rotated NS5--branes and D4--branes (see also \cite{Gubser:1998ia,Lopez:1998zf,vonUnge:1999hc} for early work on these models). The 
simplest example of this family is the SPP theory \cite{Morrison:1998cs}, which in our 
notation is $L^{1,2,1}$, and has GLSM charges $(1,-1,2,-2)$. The 
brane tiling for this theory was constructed in \cite{Franco:2005rj} and 
indeed uses one $A$ and one $C$ building block. We will see that it 
is possible to construct the entire family of gauge theories. We 
have already shown in Section \ref{section_amax} that the 
computation of R--charges and central charge using a--maximisation 
agrees with the geometric calculation for an arbitrary $L^{a,b,c}$. 
We now compute these values explicitly for this sub--family and show 
agreement with the results derived using the metric \cite{Cvetic:2005ft} and 
the toric diagram \cite{Martelli:2005tp}. These types of checks have already been 
performed for another infinite sub--family of the $L^{a,b,c}$ 
geometries, namely the $Y^{p,q}$ manifolds, in \cite{Benvenuti:2004dy} 
and \cite{Martelli:2005tp}.

\begin{figure}[ht] 
  \epsfxsize = 5cm 
  \centerline{\epsfbox{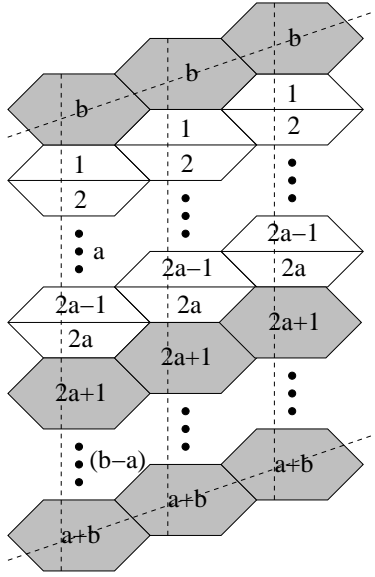}} 
  \caption{Brane tiling for $L^{a,b,a}$.} 
  \label{Laba} 
\end{figure}

\begin{figure}[ht] 
  \epsfxsize = 13cm 
  \centerline{\epsfbox{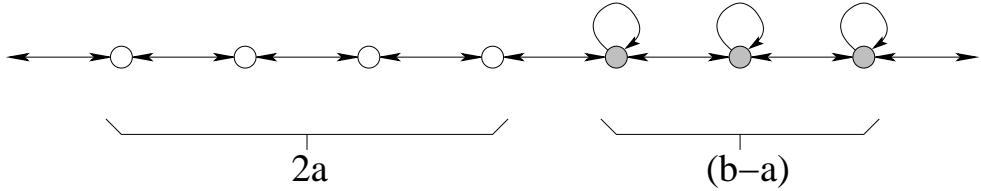}} 
  \caption{Quiver diagram for $L^{a,b,a}$. It consists of $2a$ C nodes and $(b-a)$ A nodes. The last node is connected to the 
first one by a bidirectional arrow.} 
  \label{quiver_Laba} 
\end{figure}

Let us first compute the volume of $L^{a,b,a}$ from the metric. 
The quartic equation in \cite{Cvetic:2005ft} from which the 
value of $W$ entering \eref{volume_quartic} is determined becomes 
\beq 
W^2 \left({1024 \ a^2 (a-b)^2 b^2 \over (a+b)^6}+{64 \ (2a-b)(2b-a)W \over (a+b)^2}-27 \ W^2 \right)=0~. 
\eeq 
Taking the positive solution to this equation, we obtain 
 
\beq 
\vol(L^{a,b,a})={4 \pi^3 \over 27 \ a^2 b^2} \left[(2b-a)(2a-b)(a+b)+2 \ (a^2-ab+b^2)^{3/2}\right]~. 
\label{vol_Laba} 
\eeq 
There is an alternative geometric approach to computing the volume of $L^{a,b,a}$ which uses the toric diagram instead 
of the metric: $Z$-minimisation. This method for calculating the volume of the base of a toric cone from its toric diagram was introduced 
in \cite{Martelli:2005tp}. For 3--complex dimensional cones, it corresponds to the minimisation of a two variable function $Z[y,t]$. 
The toric diagram for $L^{a,b,a}$, as we presented in Section \ref{subsection_Laba_geometry}, has vertices 
\beq [0,0] \ \ \ \ \ [1,0] \ \ \ \ \ [1,b] \ \ \ \ \ [0,a]\,. \eeq 
We then have 
\beq 
Z[y,t]=3 {y(b-a)+3a\over t(y-3)y(t-y(b-a)-3a)}~. 
\eeq 
The values of $t$ and $y$ that minimize $Z[y,t]$ are \beq 
t_{min}={1\over 2}(a+b+w) \ \ \ \ \ y_{min}={2a-b-w \over a-b} \eeq 
where \beq w=\sqrt{a^2-ab+b^2}\, . \eeq 
Computing $\vol(Y)=\pi^3 Z_{min}/3$, we recover \eref{vol_Laba}, 
which was determined using the metric. We now show how this result 
is reproduced by a gauge theory computation. The unique solution to 
\eref{AD} for the case of $L^{a,b,a}$ is 
$(n_A,n_B,n_C,n_D)=(b-a,0,a,0)$, so the brane tiling consists of 
$(b-a)$ A and $a$ C hexagons. This tiling is shown in \fref{Laba}. 
First, we note that these theories are non--chiral. \fref{quiver_Laba} shows their quiver diagram. 
 
 
Their superpotential can be easily read from the tiling in \fref{Laba}. These models do not have $V_2$ fields. Nevertheless, the parametrization of R--charges given in \eref{R-parametrization} is applicable to this case. Using it, we have 
 
\beq \tr R^3(x,y,z)={1\over 3}\left[b\left( 9 y (z-x)(x+y-z-2) -2 
\right) + a\left( 9 (2-x-y) (x+y-z) z-2\right)\right]~. 
\label{R3_Laba} \eeq Maximising \eref{R3_Laba}, we obtain the 
R--charges 
\beq 
\begin{array}{lclclcl} 
R[U_1] & = & {1 \over 3} {b-2a+w \over b-a}& \ \ \ \ \ & 
R[U_2] & = & {1 \over 3} {2b-a-w \over b-a} \\  \\ 
R[V_1] & = & {2 \over 3} {2b-a-w \over b-a} & \ \ \ \ \ & 
R[Y] & = &  {1 \over 3} {b-2a+w \over b-a} \\ \\ 
R[Z]   & = & {1 \over 3} {2b-a-w \over b-a} & \ \ \ \ \ & 
\end{array}~. 
\eeq 
The central charge $a$ is then 
 
\beq 
a(L^{aba})={27 \ a^2 b^2 \over 16}\left[(2b-a)(2a-b)(a+b)+2 \ (a^2-ab+b^2)^{3/2}\right]^{-1} 
\eeq 
which reproduces \eref{vol_Laba} on using $a=\pi^3/4 \ \vol(Y)$.

\section{Conclusions}
                                                                                                                             
The main result of this paper is the development of a combination of
techniques which allow one to extract the data defining a
(superconformal) quiver gauge theory purely from toric and
Sasaki--Einstein geometry. We have shown that the brane tiling
method provides a rather powerful organizing principle for these
theories, which generically have very intricate quivers. We emphasise
that, in the spirit of \cite{Martelli:2005tp}, our results \emph{do not} rely on
knowledge of explicit metrics, and are therefore applicable in
principle to an arbitrary toric singularity. It is nevertheless
interesting, for a variety of reasons, to know the corresponding
Sasaki--Einstein metrics in explicit form.
                                                                                                                             
For illustrating these general principles, we have discussed an
infinite family of toric singularities denoted $L^{a,b,c}$. These
generalise the $Y^{p,q}$ family, which have been the subject of much
attention; the corresponding $L^{a,b,c}$ Sasaki--Einstein metrics have been
recently constructed in \cite{Cvetic:2005ft} (see also \cite{Martelli:2005wy}). The main
input into constructing these theories came from the geometrical
data, which strongly restricts the allowed gauge theories.
Subsequently, the brane tiling technique provides a very elegant
way of organizing the data of the gauge theory. In constrast to the
quivers and superpotentials, which are very complicated to write
down in general, it is comparatively easy to describe the building
blocks of the brane tiling associated to any given $L^{a,b,c}$. We
have computed the exact R--charges of the entire family using three
different methods and found perfect agreement of the results, thus
confirming the validity of our construction.

There are several possible directions for future research suggested
by this work. One obvious question is how to extend these results to
more general toric geometries. The smooth $L^{a,b,c}$ spaces are
always described by toric diagrams with four external legs, implying
that the Sasaki--Einstein spaces (when smooth) 
have topology $S^2\times S^3$. In
fact, they are the most general such metrics. This is reflected in
the field theory by the existence of only one global baryonic $U(1)$
symmetry, giving a total of $U(1)^3$ non--R global symmetries for
these theories. But of course this is not the most general toric
geometry one might consider. One class of examples which
do not fit into the $L^{a,b,c}$ family are the $X^{p,q}$
\cite{Hanany:2005hq} spaces, whose toric diagrams have five external
legs. These include for instance the complex cone over $dP_2$, and
the Sasaki--Einstein metrics are not known explicitly. It would be
very interesting (and probably a very difficult feat) to write down
the gauge theories dual to any toric diagram. However, this may be
possible: Thanks to such advances as $Z$--minimisation
\cite{Martelli:2005tp}, and from the results of this paper, we now
know that it is possible to read off a large amount of information
from the toric data alone. In principle one could use this, along
with the brane tilings, to obtain non-trivial information about the
gauge theory dual to any toric geometry.
                                                                                                                             
One could also proceed by analogy to the progress made on the
$Y^{p,q}$ spaces. Since we know the $L^{a,b,c}$ metrics, it is
possible to seek deformations of these theories on both the
supergravity and gauge theory sides. Indeed there have been many
exciting new developments thanks to the study of the $Y^{p,q}$
theories and their cascading solutions. For the $L^{a,b,c}$ family,
the study of the supergravity side of the cascades has been initiated
in \cite{Martelli:2005wy}. Here we have not attempted to elaborate on these
deformations, but doing so would be very interesting. Additionally,
the recent progress in Sasaki-Einstein spaces opens up the question
of what other metrics we can derive for other infinite families of
spaces.
                                                                                                                             
Although the toric data specifies much information, it is still
important in many cases to know the metric. This is true for example
if one wishes to compute the Kaluza--Klein spectrum of a given
background. Another example is given by the problem of  constructing
non-conformal deformations, or cascading solutions. It will be
interesting to see whether the program initiated in \cite{Martelli:2005tp} can
be developed further, so that for instance one can extract
information on the spectrum of the Laplacian operator purely from
the toric data.

Another interesting question that arises in the context of our work
is how to prove the equivalence of $a$--maximisation and $Z$--minimisation.
Although both procedures yield the same numbers in all known examples
(of which there are infinitely many), it is nevertheless not 
{\it a priori} obvious that this should be the case. In particular,
the function one extremizes during $a$--maximisation is simply a cubic,
whereas the function used in $Z$--minimisation is a rational function.
Furthermore, $Z$--minimisation is a statement about Sasaki-Einstein spaces
in any dimension, while $a$--maximisation appears to be unique to four
dimensions. Intuitively, we suspect (thanks to AdS/CFT) that
these two procedures are equivalent, but this statement is far
from obvious. It would be wonderful to have a general proof that
$Z$--minimisation is the same as $a$--maximisation.

\section{Acknowledgements} 
 
D. M. and J. F. S. would like to thank S.--T. Yau for discussions.
S. F. and A. H. would like to thank Bo Feng, Yang-Hui He and Kris Kennaway for discussions. 
A.H. would like to thank Sergio Benvenuti for conversations.
J. F. S. would also like to thank CalTech, KITP and the University of
Cambridge for hospitality. He
is supported by NSF grants DMS--0244464, DMS--0074329 and
DMS--9803347. D. M. would like to thank his parents for warm hospitality
while this work has been completed.
B.W. thanks the University of Chicago, where some of this work 
was completed. S.F., A.H., D.V., and B.W. are supported in part
by the CTP and LNS of MIT, DOE contract $\#$DE-FC02-94ER40818, and
NSF grant PHY-00-96515. A.H. is additionally supported by the BSF
American-Israeli Bi-National Science Foundation and a DOE OJI Award.
D.V. is also supported by the MIT Praecis Presidential Fellowship.

\section{Appendix: More examples} 

In this appendix, we include additional examples that illustrate
the simplicity of our approach to the construction of brane tilings
and gauge theories. 

\subsection{Brane tiling and quiver for $L^{1,5,2}$} 
 
\begin{figure}[ht] 
  \epsfxsize = 9cm 
  \centerline{\epsfbox{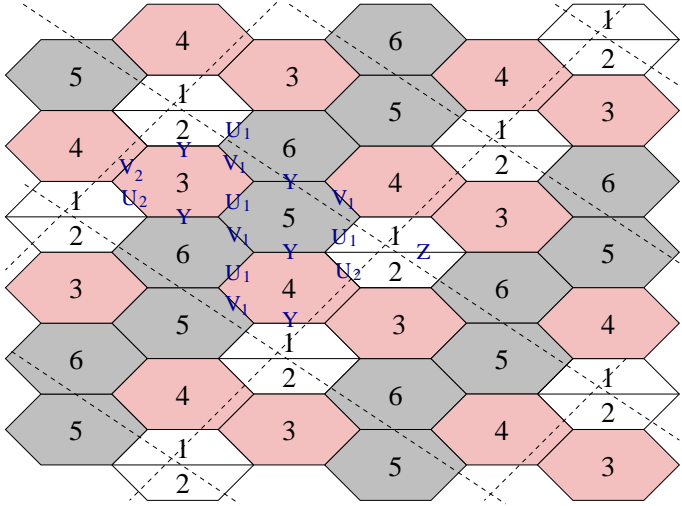}} 
  \caption{Brane tiling for $L^{1,5,2}$.} 
  \label{Y321} 
\end{figure}

\fref{Y321} shows a brane tiling for $L^{1,5,2}$ with 
$(n_A,n_B,n_C,n_D)=(2,2,1,0)$.
 
\begin{figure}[ht] 
  \epsfxsize = 7cm 
  \centerline{\epsfbox{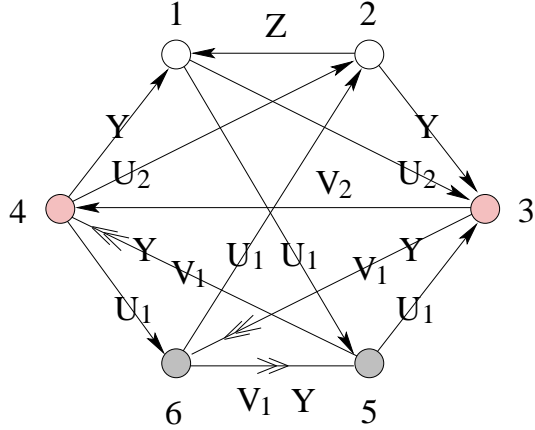}} 
  \caption{Quiver diagram for $L^{1,5,2}$.} 
  \label{quiver_L152} 
\end{figure}

The corresponding quiver diagram is shown \fref{quiver_L152}.

\begin{figure}[ht] 
  \epsfxsize = 3cm 
  \centerline{\epsfbox{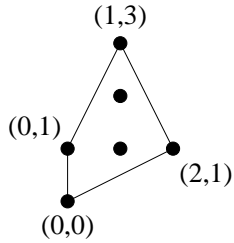}} 
  \caption{Toric diagram for $L^{1,5,2}$.} 
  \label{toric_L152} 
\end{figure}

The toric diagram computed from the tiling according to the prescription
in \cite{Hanany:2005ve,Franco:2005rj} is presented in \fref{toric_L152}.
 
\pagebreak
 
\subsection{Brane tiling and quiver for $L^{1,7,3}$} 

\begin{figure}[ht] 
  \epsfxsize = 5.5cm 
  \centerline{\epsfbox{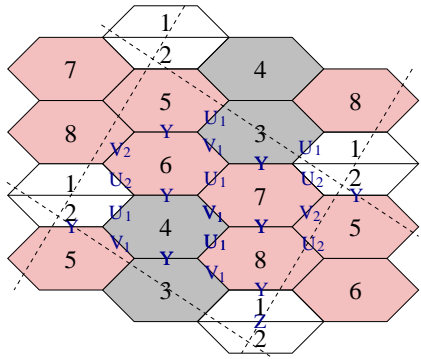}} 
  \caption{Brane tiling for $L^{1,7,3}$.} 
  \label{L173} 
\end{figure}

The brane tiling for $L^{1,7,3}$ corresponding to $(n_A,n_B,n_C,n_D)=(2,4,1,0)$
is presented in \fref{L173}.
 
\begin{figure}[ht] 
  \epsfxsize = 7cm 
  \centerline{\epsfbox{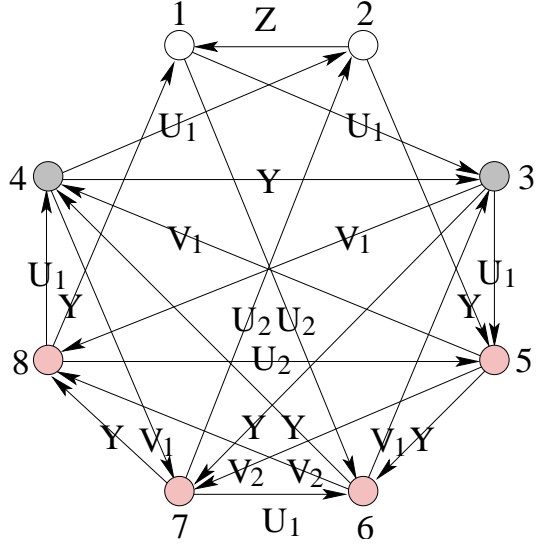}} 
  \caption{Quiver diagram for $L^{1,7,3}$.} 
  \label{quiver_L173} 
\end{figure}

\fref{quiver_L173} shows the quiver diagram for this phase.
 
\begin{figure}[ht] 
  \epsfxsize = 3cm 
  \centerline{\epsfbox{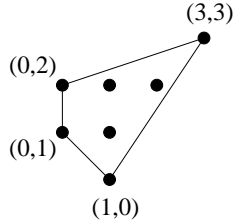}} 
  \caption{Toric diagram for $L^{1,7,3}$.} 
  \label{toric_L173} 
\end{figure}

The toric diagram is given in \fref{toric_L173}.

\bibliographystyle{JHEP} 

\end{document}

%% file: texpref.tex
\newcommand{\eref}[1]{(\ref{#1})}

\newcommand{\fref}[1]{Figure~\ref{#1}}
\newcommand{\cref}[1]{Chapter~\ref{#1}}
\newcommand{\beq}{\begin{equation}}
\newcommand{\eeq}{\end{equation}}
\newcommand{\ba}{\begin{array}}
\newcommand{\ea}{\end{array}}
\newcommand{\bcenter}{\begin{center}}
\newcommand{\ecenter}{\end{center}}

\def\IB{\relax\hbox{$\inbar\kern-.3em{\rm B}$}}
\def\IC{\relax\hbox{$\inbar\kern-.3em{\rm C}$}}
\def\ID{\relax\hbox{$\inbar\kern-.3em{\rm D}$}}
\def\IE{\relax\hbox{$\inbar\kern-.3em{\rm E}$}}
\def\IF{\relax\hbox{$\inbar\kern-.3em{\rm F}$}}
\def\IG{\relax\hbox{$\inbar\kern-.3em{\rm G}$}}
\def\IGa{\relax\hbox{${\rm I}\kern-.18em\Gamma$}}
\def\IH{\relax{\rm I\kern-.18em H}}
\def\IK{\relax{\rm I\kern-.18em K}}
\def\IL{\relax{\rm I\kern-.18em L}}
\def\IP{\relax{\rm I\kern-.18em P}}
\def\IR{\relax{\rm I\kern-.18em R}}
\def\IZ{\relax\ifmmode\mathchoice
{\hbox{\cmss Z\kern-.4em Z}}{\hbox{\cmss Z\kern-.4em Z}}
{\lower.9pt\hbox{\cmsss Z\kern-.4em Z}}
{\lower1.2pt\hbox{\cmsss Z\kern-.4em Z}}\else{\cmss Z\kern-.4em Z}\fi}
\def\II{\relax{\rm I\kern-.18em I}}

\def\sCC{{\kern 0.27em\vrule height1.45ex width0.03em depth0em
          \kern-0.30em\rm C}}
\def\C{{\mathchoice
  {\sCC}
  {\sCC}
  {\kern 0.225em \vrule height1.05ex width0.025em depth0em \kern-0.25em \rm C}
  {\kern 0.180em \vrule height0.78ex width0.02em depth0em \kern-0.2em \rm C}
        }}
\def\sHH{{\rm I\kern-.16em{}H}}
\def\H{{\mathchoice
  {\sHH}
  {\sHH}
  {\rm I\kern-.13em{}H}
  {\rm I\kern-.13em{}H} }}
\def\sNN{{\rm I\kern-.16em{}N}}
\def\N{{\mathchoice
  {\sNN}
  {\sNN}
  {\rm I\kern-.12em{}N}
  {\rm I\kern-.10em{}N} }}
\def\sPP{{\rm I\kern-.16em{}P}}
\def\P{{\mathchoice
  {\sPP}
  {\sPP}
  {\rm I\kern-.12em{}P}
  {\rm I\kern-.10em{}P} }}
\def\sQQ{{\kern 0.27em \vrule height1.45ex width0.03em depth0em
          \kern-0.30em \rm Q}}
\def\Q{{\mathchoice
        {\sQQ}
        {\sQQ}
  {\kern 0.225em \vrule height1.05ex width0.025em depth0em \kern-0.25em \rm Q}
  {\kern 0.180em \vrule height0.78ex width0.020em depth0em \kern-0.20em \rm Q}
        }}
\def\sRR{{\rm I\kern-0.16em{}R}}
\def\R{{\mathchoice
  {\sRR}
  {\sRR}
  {\rm I\kern-0.12em{}R}
  {\rm I\kern-0.10em{}R} }}
\def\sZZ{{\rm Z\kern-0.32em{}Z}}
\def\Z{{\mathchoice
  {\sZZ}
  {\sZZ} 
  {\rm Z\kern-0.3em{}Z}     
  {\rm Z\kern-0.25em{}Z} }}  
\def\ZZZ{{\rm Z\kern-0.24em{}Z}}
\def\sII{{\rm I\kern-0.16em{}I}}
\def\I{{\mathchoice
  {\sII}
  {\sII}
  {\rm I\kern-0.12em{}I}
  {\rm I\kern-0.10em{}I} }}

\def\vol{{\rm vol}}

\def\inbar{\,\vrule height1.5ex width.4pt depth0pt}
\font\cmss=cmss10 \font\cmsss=cmss10 at 7pt

\def\smiley{\hbox{\large$\bigcirc$\hspace{-0.80em}\raise.2ex
\hbox{$\cdot\cdot$}\kern-.61em\lower.2ex\hbox{\scriptsize$\smile$}}\ }
\def\frowny{\hbox{\large$\bigcirc$\hspace{-0.80em}\raise.2ex
\hbox{$\cdot\cdot$}\kern-.635em\lower.2ex\hbox{\scriptsize$\frown$}}\ }

\def\I{{\rlap{1} \hskip 1.6pt \hbox{1}}}

\makeatletter
\let\hangafter\@hangfrom
\makeatother

%% file: paper.bbl
\begin{thebibliography}{99} 
 
 
\bibitem{Maldacena:1997re}
  J.~M.~Maldacena,
  ``The large N limit of superconformal field theories and supergravity,''
  Adv.\ Theor.\ Math.\ Phys.\  {\bf 2}, 231 (1998)
  [Int.\ J.\ Theor.\ Phys.\  {\bf 38}, 1113 (1999)]
  [arXiv:hep-th/9711200].

\bibitem{Gubser:1998bc}
  S.~S.~Gubser, I.~R.~Klebanov and A.~M.~Polyakov,
  ``Gauge theory correlators from non-critical string theory,''
  Phys.\ Lett.\ B {\bf 428}, 105 (1998)
  [arXiv:hep-th/9802109].

\bibitem{Witten:1998qj}
  E.~Witten,
  ``Anti-de Sitter space and holography,''
  Adv.\ Theor.\ Math.\ Phys.\  {\bf 2}, 253 (1998)
  [arXiv:hep-th/9802150].

\bibitem{Kehagias:1998gn}
  A.~Kehagias,
  ``New Type IIB vacua and their F-theory interpretation,''
  Phys.\ Lett.\ B {\bf 435}, 337 (1998)
  [arXiv:hep-th/9805131].


\bibitem{Klebanov:1998hh}
  I.~R.~Klebanov and E.~Witten,
  ``Superconformal field theory on threebranes at a Calabi-Yau  singularity,''
  Nucl.\ Phys.\ B {\bf 536}, 199 (1998)
  [arXiv:hep-th/9807080].

\bibitem{Acharya:1998db}
  B.~S.~Acharya, J.~M.~Figueroa-O'Farrill, C.~M.~Hull and B.~Spence,
  ``Branes at conical singularities and holography,''
  Adv.\ Theor.\ Math.\ Phys.\  {\bf 2}, 1249 (1999)
  [arXiv:hep-th/9808014].

\bibitem{Morrison:1998cs}
  D.~R.~Morrison and M.~R.~Plesser,
  ``Non-spherical horizons. I,''
  Adv.\ Theor.\ Math.\ Phys.\  {\bf 3}, 1 (1999)
  [arXiv:hep-th/9810201].

\bibitem{Cvetic:2005ft}
  M.~Cvetic, H.~Lu, D.~N.~Page and C.~N.~Pope,
  ``New Einstein-Sasaki spaces in five and higher dimensions,''
  arXiv:hep-th/0504225.



\bibitem{Martelli:2005wy}
  D.~Martelli and J.~Sparks,
  ``Toric Sasaki-Einstein metrics on $S^2 \times S^3$,''
  Phys.\ Lett.\ B {\bf 621}, 208 (2005)
  [arXiv:hep-th/0505027].

\bibitem{Gauntlett:2004zh}
  J.~P.~Gauntlett, D.~Martelli, J.~Sparks and D.~Waldram,
  ``Supersymmetric AdS(5) solutions of M-theory,''
  Class.\ Quant.\ Grav.\  {\bf 21}, 4335 (2004)
  [arXiv:hep-th/0402153].

\bibitem{Gauntlett:2004yd}
  J.~P.~Gauntlett, D.~Martelli, J.~Sparks and D.~Waldram,
  ``Sasaki--Einstein metrics on $S^2 \times S^3$,''
  Adv.\ Theor.\ Math.\ Phys.\  {\bf 8}, 711 (2004)
  [arXiv:hep-th/0403002].

\bibitem{Martelli:2004wu}
  D.~Martelli and J.~Sparks,
  ``Toric geometry, Sasaki-Einstein manifolds and a new infinite class of
  AdS/CFT duals,''
  arXiv:hep-th/0411238.

\bibitem{Benvenuti:2004dy}
  S.~Benvenuti, S.~Franco, A.~Hanany, D.~Martelli and J.~Sparks,
  ``An infinite family of superconformal quiver gauge theories with
  Sasaki--Einstein duals,''
  JHEP {\bf 0506}, 064 (2005)
  [arXiv:hep-th/0411264].

\bibitem{Herzog:2004tr}
  C.~P.~Herzog, Q.~J.~Ejaz and I.~R.~Klebanov,
  ``Cascading RG flows from new Sasaki-Einstein manifolds,''
  JHEP {\bf 0502}, 009 (2005)
  [arXiv:hep-th/0412193].

\bibitem{Benvenuti:2004wx}
  S.~Benvenuti, A.~Hanany and P.~Kazakopoulos,
  ``The toric phases of the Y(p,q) quivers,''
  arXiv:hep-th/0412279.


\bibitem{Pal:2005mr}
  S.~S.~Pal,
  ``A new Ricci flat geometry,''
  Phys.\ Lett.\ B {\bf 614}, 201 (2005)
  [arXiv:hep-th/0501012].

\bibitem{Benvenuti:2005wi}
  S.~Benvenuti and A.~Hanany,
  ``Conformal manifolds for the conifold and other toric field theories,''
  arXiv:hep-th/0502043.

\bibitem{Lunin:2005jy}
  O.~Lunin and J.~Maldacena,
  ``Deforming field theories with U(1) x U(1) global symmetry and their gravity
  arXiv:hep-th/0502086.

\bibitem{Bergman:2005ba}
  A.~Bergman,
  ``Undoing orbifold quivers,''
  arXiv:hep-th/0502105.


\bibitem{Franco:2005fd}
  S.~Franco, A.~Hanany and A.~M.~Uranga,
  ``Multi-flux warped throats and cascading gauge theories,''
  arXiv:hep-th/0502113.

\bibitem{Cascales:2005rj}
  J.~F.~G.~Cascales, F.~Saad and A.~M.~Uranga,
  ``Holographic dual of the standard model on the throat,''
  arXiv:hep-th/0503079.

\bibitem{Burrington:2005zd}
  B.~A.~Burrington, J.~T.~Liu, M.~Mahato and L.~A.~Pando Zayas,
  ``Towards supergravity duals of chiral symmetry breaking in Sasaki-Einstein
  arXiv:hep-th/0504155.

\bibitem{Berenstein:2005xa}
  D.~Berenstein, C.~P.~Herzog, P.~Ouyang and S.~Pinansky,
  ``Supersymmetry breaking from a Calabi-Yau singularity,''
  arXiv:hep-th/0505029.

\bibitem{Franco:2005zu}
  S.~Franco, A.~Hanany, F.~Saad and A.~M.~Uranga,
  ``Fractional branes and dynamical supersymmetry breaking,''
  arXiv:hep-th/0505040.

\bibitem{Benvenuti:2005cz}
  S.~Benvenuti and M.~Kruczenski,
  ``Semiclassical strings in Sasaki-Einstein manifolds and long operators in N
  = 1 gauge theories,''
  arXiv:hep-th/0505046.


\bibitem{Bertolini:2005di}
  M.~Bertolini, F.~Bigazzi and A.~L.~Cotrone,
  ``Supersymmetry breaking at the end of a cascade of Seiberg dualities,''
  arXiv:hep-th/0505055.

\bibitem{Franco:2004jz}
  S.~Franco, Y.~H.~He, C.~Herzog and J.~Walcher,
  ``Chaotic duality in string theory,''
  Phys.\ Rev.\ D {\bf 70}, 046006 (2004)
  [arXiv:hep-th/0402120].


\bibitem{Klebanov:2000hb}
  I.~R.~Klebanov and M.~J.~Strassler,
  ``Supergravity and a confining gauge theory: Duality cascades and
  JHEP {\bf 0008}, 052 (2000)
  [arXiv:hep-th/0007191].

\bibitem{Hanany:2005hq}
  A.~Hanany, P.~Kazakopoulos and B.~Wecht,
  ``A new infinite class of quiver gauge theories,''
  arXiv:hep-th/0503177.

\bibitem{Ahn:2005vc}
  C.~Ahn and J.~F.~Vazquez-Poritz,
  ``Marginal Deformations with $U(1)^3$ Global Symmetry,''
  arXiv:hep-th/0505168.

\bibitem{Gauntlett:2005zz}
Jerome P.~Gauntlett, Sangmin Lee, Toni Mateos, Daniel Waldram,
``Marginal Deformations of Field Theories with $AdS_4$ Duals,''
arXiv:hep-th/0505207.



\bibitem{Intriligator:2003jj}
  K.~Intriligator and B.~Wecht,
  ``The exact superconformal R-symmetry maximizes a,''
  Nucl.\ Phys.\ B {\bf 667}, 183 (2003)
  [arXiv:hep-th/0304128].


\bibitem{Anselmi:1998am}
D.~Anselmi, D.~Z.~Freedman, M.~T.~Grisaru and A.~A.~Johansen,
``Nonperturbative formulae for central functions of supersymmetric gauge theories'',
\textsf{Nucl.~Phys.~B526,~543~(1998)},
[arXiv:hep-th/9708042].

\bibitem{Anselmi:1998ys}
D.~Anselmi, J.~Erlich, D.~Z.~Freedman and A.~A.~Johansen,
``Positivity constraints on anomalies in supersymmetric gauge
theories'', \textsf{Phys.~Rev.~D57,~7570~(1998)},
[arXiv:hep-th/9711035].

\bibitem{Henningson:1998gx}
  M.~Henningson and K.~Skenderis,
  ``The holographic Weyl anomaly,''
  JHEP {\bf 9807}, 023 (1998)
  [arXiv:hep-th/9806087].


\bibitem{Feng:2000mi}
B.~Feng, A.~Hanany and Y.~H.~He,
``D-brane gauge theories from toric singularities and toric duality,''
Nucl.\ Phys.\ B {\bf 595}, 165 (2001)
[arXiv:hep-th/0003085].

\bibitem{Feng:2001xr}
  B.~Feng, A.~Hanany and Y.~H.~He,
  ``Phase structure of D-brane gauge theories and toric duality,''
  JHEP {\bf 0108}, 040 (2001)
  [arXiv:hep-th/0104259].

\bibitem{Kutasov:2003iy}
  D.~Kutasov, A.~Parnachev and D.~A.~Sahakyan,
  ``Central charges and U(1)R symmetries in N = 1 super Yang-Mills,''
  JHEP {\bf 0311}, 013 (2003)
  [arXiv:hep-th/0308071].

\bibitem{Intriligator:2003mi}
  K.~Intriligator and B.~Wecht,
  ``RG fixed points and flows in SQCD with adjoints,''
  Nucl.\ Phys.\ B {\bf 677}, 223 (2004)
  [arXiv:hep-th/0309201].

\bibitem{Kutasov:2003ux}
  D.~Kutasov,
  ``New results on the 'a-theorem' in four dimensional supersymmetric field
  arXiv:hep-th/0312098.

\bibitem{Csaki:2004uj}
  C.~Csaki, P.~Meade and J.~Terning,
  ``A mixed phase of SUSY gauge theories from a-maximization,''
  JHEP {\bf 0404}, 040 (2004)
  [arXiv:hep-th/0403062].


\bibitem{Barnes:2004jj}
  E.~Barnes, K.~Intriligator, B.~Wecht and J.~Wright,
  ``Evidence for the strongest version of the 4d a-theorem, via a-maximization
  along RG flows,''
  Nucl.\ Phys.\ B {\bf 702}, 131 (2004)
  [arXiv:hep-th/0408156].

\bibitem{Kutasov:2004xu}
  D.~Kutasov and A.~Schwimmer,
  ``Lagrange multipliers and couplings in supersymmetric field theory,''
  Nucl.\ Phys.\ B {\bf 702}, 369 (2004)
  [arXiv:hep-th/0409029].

\bibitem{Barnes:2005zn}
  E.~Barnes, K.~Intriligator, B.~Wecht and J.~Wright,
  ``N = 1 RG flows, product groups, and a-maximization,''
  Nucl.\ Phys.\ B {\bf 716}, 33 (2005)
  [arXiv:hep-th/0502049].

\bibitem{Bertolini:2004xf}
  M.~Bertolini, F.~Bigazzi and A.~L.~Cotrone,
  ``New checks and subtleties for AdS/CFT and a-maximization,''
  JHEP {\bf 0412}, 024 (2004s
  [arXiv:hep-th/0411249].

\bibitem{Martelli:2005tp}
  D.~Martelli, J.~Sparks and S.~T.~Yau,
  ``The geometric dual of a-maximization for toric Sasaki-Einstein manifolds,''
  arXiv:hep-th/0503183.

\bibitem{Hanany:2005ve}
  A.~Hanany and K.~D.~Kennaway,
  ``Dimer models and toric diagrams,''
  arXiv:hep-th/0503149.

\bibitem{Franco:2005rj}
  S.~Franco, A.~Hanany, K.~D.~Kennaway, D.~Vegh and B.~Wecht,
  ``Brane dimers and quiver gauge theories,''
  arXiv:hep-th/0504110.

\bibitem{Benvenuti:2005ja}
  S.~Benvenuti and M.~Kruczenski,
  ``From Sasaki-Einstein spaces to quivers via BPS geodesics: Lpqr,''
  arXiv:hep-th/0505206.

\bibitem{Butti:2005sw}
  A.~Butti, D.~Forcella and A.~Zaffaroni,
  ``The dual superconformal theory for L(p,q,r) manifolds,''
  arXiv:hep-th/0505220.

\bibitem{L} E.~Lerman, ``Contact Toric Manifolds'', J. Symplectic
Geom. 1 (2003), no. 4, 785--828
arXiv:math.SG/0107201.



\bibitem{Ltop} E.~Lerman, ``Homotopy Groups of K-Contact
Toric Manifolds,''  Trans. Amer. Math. Soc.  356 (2004), no. 10,
4075--4083
math.SG/0204064.


\bibitem{Berenstein:2002ke}
D.~Berenstein, C.~P.~Herzog and I.~R.~Klebanov,
``Baryon spectra and AdS/CFT correspondence,''
JHEP {\bf 0206}, 047 (2002)
[arXiv:hep-th/0202150].

\bibitem{Intriligator:2003wr}
K.~Intriligator and B.~Wecht,
``Baryon charges in 4D superconformal field theories and their AdS duals,''
Commun.\ Math.\ Phys.\  {\bf 245}, 407 (2004)
[arXiv:hep-th/0305046].

\bibitem{Herzog:2003wt}
C.~P.~Herzog and J.~McKernan,
``Dibaryon spectroscopy,''
JHEP {\bf 0308}, 054 (2003)
[arXiv:hep-th/0305048].

\bibitem{Herzog:2003dj}
C.~P.~Herzog and J.~Walcher,
``Dibaryons from exceptional collections,''
JHEP {\bf 0309}, 060 (2003)
[arXiv:hep-th/0306298].


\bibitem{Aharony:1997bh}
O.~Aharony, A.~Hanany and B.~Kol,
``Webs of (p,q) 5-branes, five dimensional field theories and grid
diagrams,''
JHEP {\bf 9801}, 002 (1998)
[arXiv:hep-th/9710116].

\bibitem{Aharony:1997ju}
O.~Aharony and A.~Hanany,
``Branes, superpotentials and superconformal fixed points,''
Nucl.\ Phys.\ B {\bf 504}, 239 (1997)
[arXiv:hep-th/9704170].

\bibitem{Leung:1997tw}
N.~C.~Leung and C.~Vafa,
``Branes and toric geometry,''
Adv.\ Theor.\ Math.\ Phys.\  {\bf 2}, 91 (1998)
[arXiv:hep-th/9711013].

\bibitem{Hanany:2001py}
A.~Hanany and A.~Iqbal,
``Quiver theories from D6-branes via mirror symmetry,''
JHEP {\bf 0204}, 009 (2002)
[arXiv:hep-th/0108137].

\bibitem{LT} 
E.~Lerman, S.~Tolman, 
"Symplectic Toric Orbifolds", 
arXiv: dg-ga/9412005

\bibitem{Feng:2002zw}
B.~Feng, S.~Franco, A.~Hanany and Y.~H.~He,
``Symmetries of toric duality,''
JHEP {\bf 0212}, 076 (2002)
[arXiv:hep-th/0205144].

\bibitem{Seiberg:1994pq}
  N.~Seiberg,
  ``Electric - magnetic duality in supersymmetric nonAbelian gauge theories,''
  Nucl.\ Phys.\ B {\bf 435}, 129 (1995)
  [arXiv:hep-th/9411149].

\bibitem{Uranga:1998vf}
  A.~M.~Uranga,
  ``Brane configurations for branes at conifolds,''
  JHEP {\bf 9901}, 022 (1999)
  [arXiv:hep-th/9811004].

\bibitem{Gubser:1998ia}
  S.~Gubser, N.~Nekrasov and S.~Shatashvili,
  ``Generalized conifolds and four dimensional N = 1 superconformal
  theories,''
  JHEP {\bf 9905}, 003 (1999)
  [arXiv:hep-th/9811230].

\bibitem{Lopez:1998zf}
  E.~Lopez,
  ``A family of N = 1 SU(N)**k theories from branes at singularities,''
  JHEP {\bf 9902}, 019 (1999)
  [arXiv:hep-th/9812025].

\bibitem{vonUnge:1999hc}
  R.~von Unge,
  ``Branes at generalized conifolds and toric geometry,''
  JHEP {\bf 9902}, 023 (1999)
  [arXiv:hep-th/9901091].

\end{thebibliography}
